# Fast Approximation Algorithms for Cut-based Problems in Undirected Graphs

Aleksander Mądry*
Massachusetts Institute of Technology
madry@mit.edu

**Abstract**

We present a general method of designing fast approximation algorithms for cut-based minimization problems in undirected graphs. In particular, we develop a technique that given any such problem that can be approximated quickly on trees, allows approximating it almost as quickly on general graphs while only losing a poly-logarithmic factor in the approximation guarantee.

To illustrate the applicability of our paradigm, we focus our attention on the undirected sparsest cut problem with general demands and the balanced separator problem. By a simple use of our framework, we obtain poly-logarithmic approximation algorithms for these problems that run in time close to linear.

The main tool behind our result is an efficient procedure that decomposes general graphs into simpler ones while approximately preserving the cut-flow structure. This decomposition is inspired by the cut-based graph decomposition of Räcke that was developed in the context of oblivious routing schemes, as well as, by the construction of the ultrasparsifiers due to Spielman and Teng that was employed to preconditioning symmetric diagonally-dominant matrices.

## 1 Introduction

Cut-based graph problems are ubiquitous in optimization. They have been extensively studied – both from theoretical and applied perspective – in the context of flow problems, theory of Markov chains, geometric embeddings, clustering, community detection, VLSI layout design, and as a basic primitive for divide-and-conquer approaches (cf. [38]). For the sake of concreteness, in this paper we define an optimization problem $\mathcal{P}$ to be an *(undirected) cut-based (minimization) problem* if its every instance $P \in \mathcal{P}$ can be cast as a task of finding – for a given input undirected graph $G = (V, E, u)$ – a cut $C^*$ such that:

$$C^* = \mathrm{argmin}_{\emptyset \neq C \subset V} u(C) f_P(C), \qquad (1)$$

where $u(C)$ is the capacity of the cut $C$ in $G$ and $f_P$ is a non-negative function that depends on $P$, but not on the graph $G$. It is not hard to see that a large number of graph problems fits this definition. For illustrative purposes, in this paper we focus on two of them that are fundamental in graph partitioning: the generalized sparsest cut problem and the balanced separator problem.

---
*Supported by NSF contract CCF-0829878 and by ONR grant N00014-05-1-0148.

In the *generalized sparsest cut* problem, we are given a graph $G = (V, E, u)$ and a demand graph $D = (V, E_D, d)$. Our task is to find a cut $C^*$ that minimizes the *(generalized) sparsity* $u(C)/d(C)$ among all the cuts $C$ of $G$. Here, $d(C)$ is the total demand $\sum_{e \in E_D(C)} d(e)$ of the demand edges $E_D(C)$ of $D$ cut by the cut $C$. Casted in our terminology, the problem is to find a cut $C^*$ being $\operatorname{argmin}_C u(C) f_P(C)$ with $f_P(C) := 1/d(C)$.

An important problem that is related to the generalized sparsest cut problem is the *sparsest cut* problem. In this problem one aims at finding a cut $C^*$ that minimizes the *sparsity* $u(C)/\min\{|C|, |\overline{C}|\}$ among all the cuts $C$ of $G$. Note that if we considered an instance of generalized sparsest cut problem corresponding to the demand graph $D$ being complete and each demand being $1/|V|$ (this special case is sometimes called the *uniform sparsest cut* problem) then the corresponding generalized sparsity of every cut $C$ is within factor of two of its sparsity. Therefore, up to this constant factor, one can view the sparsest cut problem as a special case of the generalized sparsest cut problem.

The *balanced separator* problem corresponds to a task of finding a cut that minimizes the sparsity among all the *c-balanced cuts* $C$ in $G$, i.e. among all the cuts $C$ with $\min\{|C|, |\overline{C}|\} \geq c|V|$, for some specified constant $c > 0$ called the *balance constant* of the instance. In our framework, the function $f_P(C)$ corresponding to this problem is equal to $1/\min\{|C|, |\overline{C}|\}$ if $\min\{|C|, |\overline{C}|\} \geq c|V|$ and equal to $+\infty$ otherwise.

## 1.1 Previous Work on Graph Partitioning

The captured by the above-mentioned problems task of partitioning a graph into relatively large pieces while minimizing the number of edges cut is a fundamental combinatorial problem (see [38]) with a long history of theoretical and applied investigation. Since most of the graph partitioning problems – in particular, the ones we consider – are NP-hard, one needs to settle for approximation algorithms. In case of the sparsest cut problem and the balanced separator problem, there were two main types of approaches to approximating them.

First type of approach was based on spectral methods. Inspired by Cheeger's inequality [17] discovered in the context of Riemannian manifolds, Alon and Milman [4] transplanted it to discrete setting and presented an algorithm for the sparsest cut problem that can be implemented to run in nearly-linear time[1], but its approximation ratio depends on the conductance[2] $\Phi(G)$ of the graph and can be as large as $\Omega(n)$ in the worst case. Later, Spielman and Teng [40, 42] (see also [5] and [6] for a follow-up work) used this approach to design an algorithm for the balanced separator problem. However, both the running time and the approximation guarantee still depend on the conductance of the graph, leading to an $\Omega(n)$ approximation ratio and large running time in the worst case.

The second type of approach relies on flow methods. It was started by Leighton and Rao [32]

---

[1]The key algorithmic step of [4] is computation of a vector that is orthogonal to all-ones vector and (up to sufficient precision) minimizes the energy of the Laplacian of the graph. The canonical approach to this task leads to an algorithm running in $\widetilde{O}(m/\Phi^2)$ time. To get no dependence on $\Phi$ in the running time, one notes that finding a vector that, say, 1/2-approximately *maximizes* the energy of the *pseudo-inverse* of the Laplacian is sufficient for [4] to work. One can compute such a vector in $\widetilde{O}(m)$ time by employing the algorithm of [31] and avoiding explicit computation of the pseudo-inverse via appropriate use of the nearly-linear Laplacian system solver of Spielman and Teng [40].

[2] *Conductance* $\Phi(G)$ of a graph $G = (V, E, u)$ is defined as $\Phi(G) := \min_{\emptyset \neq C \subset V} \frac{u(C)}{U(C)U(\overline{C})}$, where $U(C) := \sum_{(v,w) \in E, v \in C} u((v,w))$ is the total capacity-weighted degree of the vertices in $C$.

who used linear programming relaxation of the maximum concurrent flow problem to establish first poly-logarithmic approximation algorithms for many graph partitioning problems. In particular, an $O(\log n)$-approximation for the sparsest cut and the balanced separator problems. Both these algorithms can be implemented to run in $\widetilde{O}(n^2)$ time.

These spectral and flow-based approaches were combined in a seminal result of Arora, Rao, and Vazirani [10] to give an $O(\sqrt{\log n})$-approximation for the sparsest cut and the balanced separator problems.

In case of the generalized sparsest cut problem, the results of Linial, London, and Rabinovich [33], and of Aumann and Rabani [11] give a $O(\log r)$-approximation algorithms, where $r$ is the number of vertices of the demand graph that are endpoints of some demand edge. Subsequently, Chawla, Gupta, and Räcke [16] extended the techniques from [10] to obtain an $O(\log^{3/4} r)$-approximation. This approximation ratio was later improved by Arora, Lee, and Naor [9] to $O(\sqrt{\log r} \log \log r)$.

## 1.2 Fast Approximation Algorithms for Graph Partitioning Problems

More recently, a lot of effort was put into designing algorithms for the graph partitioning problems that are very efficient while still having poly-logarithmic approximation guarantee. Arora, Hazan, and Kale [7] combined the concept of expander flows that was introduced in [10] together with multicommodity flow computations to obtain $O(\sqrt{\log n})$-approximation algorithms for the sparsest cut and the balanced separator problems that run in $\widetilde{O}(n^2)$ time.

Subsequently, Khandekar, Rao, and Vazirani [30] designed a primal-dual framework for graph partitioning algorithms and used it to achieve $O(\log^2 n)$-approximations for the same two problems running in $\widetilde{O}(m + n^{3/2})$ time needed to perform graph sparsification of Benczúr and Karger [14] followed by a poly-logarithmic number of maximum single-commodity flow computations. In [8], Arora and Kale introduced a general approach to approximately solving semi-definite programs which, in particular, led to $O(\log n)$-approximation algorithms for the sparsest cut and the balanced separator problems that also run in $\widetilde{O}(m + n^{3/2})$ time. Later, Orecchia, Schulman, Vazirani, and Vishnoi [34] obtained the same approximation ratio and running time as [8] by extending the framework from [30]. Recently, Sherman [37] presented an algorithm that for any $\varepsilon > 0$ works in $\widetilde{O}(m + n^{3/2+\varepsilon})$ time – corresponding to performing sparsification and $\widetilde{O}(n^\varepsilon)$ maximum single-commodity flow computations – and achieves an approximation ratio of $O(\sqrt{\log n/\varepsilon})$. Therefore, for any fixed $\varepsilon > 0$, his algorithm has the best-known approximation guarantee while still having running time close to the time needed for maximum single-commodity flow computation.

Unfortunately, despite this substantial progress in designing efficient poly-logarithmic approximation algorithms for the sparsest cut problem, if one is interested in the generalized sparsest cut problem then a folklore $O(\log r)$-approximation algorithm running in $\widetilde{O}(n^2 \log U)$ time is still the fastest one known. This folklore result is obtained by combining an efficient implementation [14, 24] of the algorithm of Leighton and Rao [32] together with the results of Linial, London, and Rabinovich [33], and of Aumann and Rabani [11].

## 1.3 Our Contribution

In this paper we present a general method of designing fast approximation algorithms for undirected cut-based minimization problems.[3] In particular, we design a procedure that given an integer

---
[3]In fact, as we briefly mention in section 4.4, our framework can be applied to even more general class of problems: the multicut-based minimization problems.

**(Uniform) sparsest cut and balanced separator problem:**

| Algorithm | Approximation ratio | Running time |
|---|---|---|
| Alon-Milman [4] | $O(\Phi^{-1/2})$ ($\Omega(n)$ worst-case) | $\widetilde{O}(m/\Phi^2)$ ($\widetilde{O}(m)$ using [40]) |
| Andersen-Peres [6] | $\widetilde{O}(\Phi^{-1/2})$ ($\widetilde{\Omega}(n)$ worst-case) | $\widetilde{O}(m/\Phi^{3/2})$ |
| Leighton-Rao [32] | $O(\log n)$ | $\widetilde{O}(n^2)$ |
| Arora-Rao-Vazirani [10] | $O(\sqrt{\log n})$ | polynomial time |
| Arora-Hazan-Kale [7] | $O(\sqrt{\log n})$ | $\widetilde{O}(n^2)$ |
| Khandekar-Rao-Vazirani [30] | $O(\log^2 n)$ | $\widetilde{O}(m + n^{3/2})$ |
| Arora-Kale [8] | $O(\log n)$ | $\widetilde{O}(m + n^{3/2})$ |
| Orecchia-Schulman-Vazirani-Vishnoi [34] | $O(\log n)$ | $\widetilde{O}(m + n^{3/2})$ |
| Sherman [37] | $O(\sqrt{\log n/\varepsilon})$ | $\widetilde{O}(m + n^{3/2+\varepsilon})$ |
| this paper $k = 1$ | $(\log^{3/2+o(1)} n)/\sqrt{\varepsilon}$ | $\widetilde{O}(m + n^{6/5+\varepsilon})$ |
| this paper $k = 2$ | $(\log^{5/2+o(1)} n)/\sqrt{\varepsilon}$ | $\widetilde{O}(m + n^{12/11+\varepsilon})$ |
| this paper $k \geq 1$ | $(\log^{(1+o(1))(k+1/2)} n)/\sqrt{\varepsilon}$ | $\widetilde{O}(m + 2^k n^{1+1/(3 \cdot 2^k - 1)+\varepsilon})$ |

**Generalized sparsest cut problem:**

| Algorithm | Approximation ratio | Running time |
|---|---|---|
| Folklore ([32, 33, 11, 14, 24]) | $O(\log r)$ | $\widetilde{O}(n^2 \log U)$ |
| Chawla-Gupta-Räcke [16] | $O(\log^{3/4} r)$ | polynomial time |
| Arora-Lee-Naor [9] | $O(\sqrt{\log r} \log \log r)$ | polynomial time |
| this paper $k = 2$ | $\log^{2+o(1)} n$ | $\widetilde{O}(m + |E_D| + n^{4/3} \log U)$ |
| this paper $k = 3$ | $\log^{3+o(1)} n$ | $\widetilde{O}(m + |E_D| + n^{8/7} \log U)$ |
| this paper $k \geq 1$ | $\log^{(1+o(1))k} n$ | $\widetilde{O}(m + |E_D| + 2^k n^{1+1/(2^k-1)} \log U)$ |

Figure 1: Here, $n$ denotes the number of vertices of the input graph $G$, $m$ the number of its edges, $\Phi$ its conductance, $U$ is its capacity ratio, and $r$ is the number of vertices of the demand graph $D = (V, E_D, d)$ that are endpoints of some demand edges. Also, $\widetilde{O}(\cdot)$ notation suppresses polylogarithmic factors. The algorithm of Alon and Milman applies only to the sparsest cut problem.

parameter $k \geq 1$ and any undirected graph $G$ with $m$ edges, $n$ vertices, and having integral edge capacities in the range $[1, \ldots, U]$, produces in $\widetilde{O}(m + 2^k n^{1+1/(2^k-1)} \log U)$ time a small number of trees $\{T_i\}_i$. These trees have a property that, with high probability, for any $\alpha \geq 1$, we can find an $(\alpha \log^{(1+o(1))k} n)$-approximation to given instance of *any* cut-based minimization problem on $G$ by just obtaining some $\alpha$-optimal solution for each $T_i$ and choosing the one among them that leads to the smallest objective value in $G$ (see Theorem 4.1 for more details). As a consequence, we are able to transform any $\alpha$-approximation algorithm for a cut-based problem that works only on tree instances, to an $(\alpha \log^{(1+o(1))k} n)$-approximation algorithm for general graphs, while paying a computational overhead of $\widetilde{O}(m + 2^k n^{1+1/(2^k-1)} \log U)$ time that, as $k$ grows, quickly becomes close to linear.

We illustrate the applicability of our paradigm on two fundamental graph partitioning problems: the undirected (generalized) sparsest cut and the balanced separator problems. By a simple use of our framework we obtain, for any integral $k \geq 1$, a $(\log^{(1+o(1))k} n)$-approximation algorithm for the generalized sparsest cut problem that runs in time $\widetilde{O}(m + |E_D| + 2^k n^{1+1/(2^k-1)} \log U)$, where $|E_D|$ is the number of the demand edges in the demand graph. Furthermore, in case of the sparsest cut and the balanced separator problems, we combine our techniques with the algorithm of Sherman [37] to obtain approximation algorithms that have even better dependence on $k$ in the running time. Namely, for any $k \geq 1$ and $\varepsilon > 0$, we present $(\log^{(1+o(1))(k+1/2)} n/\sqrt{\varepsilon})$-approximation algorithms[4] for these problems that run in time $\widetilde{O}(m + 2^k n^{1+1/(3 \cdot 2^k - 1) + \varepsilon})$. We summarize our results together with the previous work in Figure 1. One can see that even for small values of $k$, the running times of our algorithms beat the multicommodity flow barrier of $\Omega(n^2)$ and the bound of $\Omega(m + n^{3/2})$ time corresponding to performing sparsification and single-commodity flow computation. Unfortunately, in the case of the generalized sparsest cut problem our approximation guarantee has poly-logarithmic dependence on $n$. This is in contrast to the previous algorithms for this problem that have their approximation guarantee depending on $r$.

## 1.4 Overview of Our Techniques

Our approach is inspired by the cut-based graph decomposition of Räcke [36] that was developed in the context of oblivious routing schemes (see [35, 12, 15, 28] for some previous work on this subject). This decomposition is a powerful tool in designing approximation algorithms for various undirected cut-based problem. It is based on finding for a given graph $G$ with $n$ vertices and $m$ edges, a convex combination $\{\lambda_i\}_i$ of decomposition trees $\{T_i\}_i$ such that: $G$ is embeddable into each $T_i$, and this convex combination can be embedded into $G$ with $O(\log n)$ congestion. One employs this decomposition by first approximating the desired cut-based problem on each tree $T_i$ – this usually yields much better approximation ratio than general instances – and then extracting from obtained solutions a solution for the graph $G$ while incurring only additional $O(\log n)$ factor in the approximation guarantee.

The key question motivating our results is: can the above paradigm be applied to obtain *very efficient* approximation algorithms for cut-based graph problems? After all, one could envision a generic approach to designing fast approximation algorithms for such problems in which one decomposes first an input graph $G$ into a convex combination of structurally simple graphs $G_i$ (e.g. trees), solves the problem quickly on each of these easy instances and then combines these solutions to obtain a solution for the graph $G$, while losing some (e.g. poly-logarithmic) factor in approximation guarantee as a price of this speed-up. Clearly, the viability of such scenario depends critically on how fast a suitable decomposition can be computed and how 'easy' are the graphs $G_i$s from the point of view of the problems we want to solve.

We start investigation of this approach by noticing that if one is willing to settle for approximation algorithms that are of Monte Carlo-type then, given a decomposition of the graph $G$ into a convex combination $\{(\lambda_i, G_i)\}_i$, one does not need to compute the solution for each of $G_i$s to get the solution for $G$. It is sufficient to just solve the problem on a small number of $G_i$s that are sampled from the distribution described by $\lambda_i$s.

However, even after making this observation, one still needs to compute the decomposition of the graph to be able to sample from it and, unfortunately, Räcke's decomposition – that was aimed at obtaining algorithms that are just polynomial-time – has serious limitations when one is interested in time-efficiency. In particular, the running time of his decomposition procedure is dominated by $\widetilde{O}(m)$ all-pair shortest path computations and is thus prohibitively large from our point of view – see section 3.1 for more details.

---

[4]As was the case in all the previous work we described, we only obtain a pseudo-approximation for the balanced separator problem. Recall that an $\alpha$-*pseudo-approximation* for the balanced separator problem with balance constant $c$ is a $c'$-balanced cut $C$ – for some other constant $c'$ – such that $C$'s sparsity is within $\alpha$ of the sparsest $c$-balanced cut. For the sake of convenience, in the rest of the paper we don't differentiate between pseudo- and true approximation.

To circumvent this problem, we design an alternative and more general graph decomposition (see Theorem 3.6). Similarly to the case of Räcke's decomposition, our construction is based on embedding graph metrics into tree metrics. However – instead of the embedding result of Fakcharoenphol, Talwar, and Rao [23] that was used by Räcke – we employ the nearly-linear time algorithm of Abraham, Bartal, and Neiman [1] for finding low-average-stretch spanning trees. This choice allows for a much more efficient implementation of our decomposition procedure at a cost of decreasing the quality of provided approximation from $O(\log n)$ to $\widetilde{O}(\log n)$. Also, even more crucially, inspired by the notion of the ultrasparsifiers of Spielman and Teng [40, 41], we allow flexibility in choosing the type of graphs into which our graph is decomposed. In this way, we are able to offer a trade-off between the structural simplicity of these graphs and the time needed to compute the corresponding decomposition.

Finally, by recursive use of our decomposition – together with sparsification technique – we are able to leverage the above-mentioned flexibility to design a procedure that can have its running time be arbitrarily close to nearly-linear and, informally speaking, allows us to sample from a decomposition of our input graph $G$ into graphs whose structure is arbitrarily simple, but at a cost of getting proportionally worse quality of the reflection of the cut structure of $G$ – see Theorem 3.7 for more details.

## 1.5 Outline of the Paper

We start with some preliminaries in section 2. Next, in section 3, we introduce the key concepts of our paper and state the main theorems. In section 4 we show how our framework leads to fast approximation algorithms for undirected cut-based minimization problems. We also construct there our algorithms for the (generalized) sparsest cut and balanced separator problems and remark on extending our framework to make it handle multicut-based minimization problems. Section 5 contains the description of our decomposition procedure that allows expressing general graphs as a convex combinations of simpler ones while approximately preserving cut-flow structure. We conclude in section 6 with showing how a recursive use of this decomposition procedure together with sparsification leads to the sampling procedure underlying our framework.

## 2 Preliminaries

In this paper we will be concerned with an undirected graphs $G = (V, E, u)$ having vertex set $V$, edge set $E$, and integer capacities $u : E \to \mathbb{Z}^+$ on the edges. By *capacity ratio* $U$ *of* $G$ we mean the maximum possible ratio between capacities of two edges of $G$ i.e. $U := \max_{e,e' \in E} u(e)/u(e')$.

For a given graph $G = (V, E, u)$, by a *cut* $C$ in $G$ we mean any vertex set $\emptyset \neq C \subset V$. We denote by $E(C)$ the set of all the edges of $G$ with exactly one endpoint in $C$ – we say that these edges are *cut* by $C$. By $\overline{C}$ we will denote the set $V \setminus C$. Also, we define the *capacity* $u(C)$ of a cut in $G$ to be the total capacity $u(E(C))$ of all the edges in $E(C)$. Finally, for some $V' \subseteq V$, we say that a subgraph $G' = (V', E', u')$ is a *subgraph of* $G$ *induced by* $V'$ if $E'$ consists of all the edges of $G$ whose both endpoints are in $V'$ and the capacity function $u'$ on these edges is inherited from the capacity function $u$ of $G$.

## 2.1 Maximum Concurrent Flow Problem

For a given graph $G = (V, E, u)$, by a *multicommodity flow* $f = (f^1, \ldots, f^k)$ we mean a set of $k$ flows $f^i$ in $G$, where each flow $f^i$ routes a commodity $i$ from some *source* $s_i$ to some *sink* $t_i$. Let us also define $|f(e)|$ as the total flow routed along the edge $e$ by $f$, i.e. $|f(e)| := \sum_i |f^i(e)|$. We say that a multicommodity flow $f$ is *feasible* if for every edge $e$, $|f(e)| \leq u(e)$.

One of the most popular multicommodity flow problems is the *maximum concurrent flow* problem. In this problem, in addition to the graph $G = (V, E, u)$, we are given a *demand graph* $D = (V, E_D, d)$. The objective is to find a feasible multicommodity flow $f$ in $G$ such that for each demand edge $e = (v, w) \in E_D$ there is a flow $f^e$ that routes $\theta d(e)$ units of corresponding commodity from $v$ to $w$ and the *flow rate* $\theta$ is maximized. Another important flow problem – that can be viewed as a special case of the maximum concurrent flow problem – is the *maximum s-t flow* problem (or just the *maximum flow* problem) that corresponds to a task of finding a feasible single-commodity flow that maximizes the *throughput* being the amount of flow pushed from the source $s$ to the sink $t$.

## 2.2 Embedability

A notion we will be dealing extensively with is the notion of graph embedding.

**Definition 2.1.** *For given graphs $G = (V, E, u)$ and $\overline{G} = (V, \overline{E}, \overline{u})$, by an embedding $f$ of $G$ into $\overline{G}$ we mean a multicommodity flow $f = (f^{e_1}, \ldots, f^{e_{|E|}})$ in $\overline{G}$ with $|E|$ commodities indexed by the edges of $G$, such that for each $1 \leq i \leq |E|$, the flow $f^{e_i}$ routes in $\overline{G}$ $u(e_i)$ units of flow between endpoints of $e_i$.*

One may view an embedding $f$ of $G$ into $\overline{G}$ as a concurrent flow in $\overline{G}$ whose source-sink pairs and corresponding demands are given by the endpoints of edges of $G$ and their corresponding capacities. Note that the above definition does not require that $f$ is feasible. We proceed to the definition of embedability.

**Definition 2.2.** *For given $t \geq 1$, graphs $G = (V, E, u)$, and $\overline{G} = (V, \overline{E}, \overline{u})$, we say that $G$ is $t$-embeddable into $\overline{G}$ if there exists an embedding $f$ of $G$ into $\overline{G}$ such that for all $e \in \overline{E}$, $|f(e)| \leq t\overline{u}(e)$. We say that $G$ is embeddable into $\overline{G}$ if $G$ is $1$-embeddable into $\overline{G}$.*

Intuitively, the fact that $G$ is $t$-embeddable into $\overline{G}$ means that we can fractionally pack all the edges of $G$ into $\overline{G}$ with all its capacities $\overline{u}$ multiplied by $t$. Also, it is easy to see that for given graphs $G$ and $\overline{G}$, one can always find the smallest possible $t$ such that $G$ is $t$-embeddable into $\overline{G}$ by just solving maximum concurrent flow problem in $\overline{G}$ in which we treat $G$ as a demand graph with the demands given by its capacities.

## 2.3 Graph Sparsification

A powerful tool in designing fast algorithms for cut-related problems is *graph sparsification* – this procedure allows approximation of all the cuts of any – possibly dense – graph by its sparse subgraph with appropriately chosen capacities. In particular, the following theorem was proved by Benczúr and Karger [14] (see also [43], [39], and [13] for more general results on sparsification).

**Theorem 2.3** ([14]). *Given graph $G = (V, E, u)$ and an accuracy parameter $\delta > 0$, there is a Monte Carlo algorithm that finds in $\widetilde{O}(|E|)$ time a subgraph $\widetilde{G} = (V, \widetilde{E}, \widetilde{u})$ of $G$ such that*

(i) $\widetilde{G}$ has $O(|V|\log|V|\delta^{-2})$ edges;

(ii) for any cut $\emptyset \neq C \subset V$ of $G$, we have $u(C) \leq \widetilde{u}(C) \leq (1+\delta)u(C)$ i.e. the cuts of $G$ are $(1+\delta)$-approximately preserved in $\widetilde{G}$;

(iii) the capacity ratio of $\widetilde{G}$ is at most $O(|V|)$ times the capacity ratio of $G$.

## 3 Graph Decompositions and Fast Approximation Algorithms for Cut-based Problems

The central concept of our paper is the notion of $(\alpha, \mathcal{G})$-decomposition. This notion is a generalization of the cut-based graph decomposition introduced by Räcke [36].

**Definition 3.1.** *For any $\alpha \geq 1$ and some family of graphs $\mathcal{G}$, by an $(\alpha, \mathcal{G})$-decomposition of a graph $G = (V, E, u)$ we mean a set of pairs $\{(\lambda_i, G_i)\}_i$, where for each $i$, $\lambda_i > 0$ and $G_i = (V, E, u_i)$ is a graph in $\mathcal{G}$, such that:*

(a) $\sum_i \lambda_i = 1$

(b) $G$ is embeddable into each $G_i$

(c) There exist embeddings $f_i$ of each $G_i$ into $G$ such that for each $e \in E$, $\sum_i \lambda_i |f_i(e)| \leq \alpha u(e)$.

*Moreover, we say that such decomposition is $k$-sparse if it consists of at most $k$ different $G_i$.*

The crucial property of $(\alpha, \mathcal{G})$-decomposition is that it captures $\alpha$-approximately the cut structure of the graph $G$. We formalize this statement in the following easy to prove fact.

**Fact 3.2.** *For any $\alpha \geq 1$, graph $G = (V, E, u)$, and a family of graphs $\mathcal{G}$, if $\{(\lambda_i, G_i)\}_i$ is an $(\alpha, \mathcal{G})$-decomposition of $G$ then for any cut $C$ of $G$:*

**(lowerbounding)** *for all $i$, the capacity $u_i(C)$ of $C$ in $G_i$ is at least its capacity $u(C)$ in $G$;*

**(upperbounding in expectation)** $\mathbb{E}_{\vec{\lambda}}[u(C)] := \sum_i \lambda_i u_i(C) \leq \alpha u(C)$.

Note that the condition (a) from Definition 3.1 implies that $\{(\lambda_i, G_i)\}_i$ is a convex combination of the graphs $\{G_i\}_i$. Therefore, one of the implications of the Fact 3.2 is that for any cut $C$ in $G$, not only the capacity of $C$ in every $G_i$ is lowerbounded by its capacity $u(C)$ in $G$, but also there always exists a graph $G_j$ in which the capacity of $C$ is at most $\alpha u(C)$. As one can easily convince oneself, for any $\beta \geq 1$, this property alone allows to reduce a task of $\alpha\beta$-approximation of any cut-based minimization problem in $G$ to a task of $\beta$-approximating this problem in each of $G_i$s.

In fact, if one is willing to settle for Monte Carlo-type approximation guarantees, one can reduce the task of approximation of such problem to a task of approximating it in only a *small number* of $G_i$s. To make this statement precise, let us introduce the following definition.

**Definition 3.3.** *For given $G = (V, E, u)$, $\alpha \geq 1$, and $1 \geq p > 0$, we say that a collection $\{G_i\}_i$ of random graphs[5] $G_i = (V, E_i, u_i)$, $\alpha$-preserves the cuts of $G$ with probability $p$ if for every cut $C$ of $G$:*

---

[5] We formalize the notion of random graphs by viewing each $G_i$ as a graph on vertex set $V$ chosen according to some underlying distribution over all such graphs.

**(lowerbounding)** *for all $i$, the capacity $u_i(C)$ of $C$ in $G_i$ is at least its capacity $u(C)$ in $G$;*

**(probabilistic upperbounding)** *with probability at least $p$, there exists $i$ such that $u_i(C) \leq \alpha u(C)$.*

With a slight abuse of notation, we will say that a random graph $G'$ is $\alpha$-preserving the cuts of $G$ with probability $p$ if the corresponding singleton family $\{G'\}$ is doing it. Now, a useful connection between $(\alpha, \mathcal{G})$-decomposition of graph $G$ and obtaining graphs $O(\alpha)$-preserving the cuts of $G$ with some probability is given by the following fact whose proof is a straight-forward application of Markov's inequality.

**Fact 3.4.** *Let $G = (V, E, u)$, $\alpha \geq 1$, $1 > p > 0$ and let $\{(\lambda_i, G_i)\}_i$ be some $(\alpha, \mathcal{G})$-decomposition of $G$. If $G'$ is a random graph chosen from the set $\{G_i\}_i$ according to distribution given by $\lambda_i$s i.e. $Pr[G' = G_i] = \lambda_i$, then $G'$ $2\alpha$-preserves cuts of $G$ with probability $1/2$.*

Therefore – as we will formally prove later – for any $\beta \geq 1$, we can obtain a Monte Carlo $2\alpha\beta$-approximation algorithm for any minimization cut-based problem in $G$, by $\beta$-approximating it in a sample of $O(\ln |V|)$ $G_i$s that were chosen according to distribution described by $\lambda_i$s.

## 3.1 Finding a Good $(\alpha, \mathcal{G})$-decomposition Efficiently

In the light of the above discussion, we focus our attention on the task of developing a $(\alpha, \mathcal{G})$-decomposition of graphs into some family $\mathcal{G}$ of structurally simple graphs that enjoys relatively small value of $\alpha$. Of course, since our emphasis is on obtaining algorithms that are very efficient, an important aspect of the decomposition that we will be interested in is the time needed to compute it.

With this goal in mind, we start by considering the following theorem due to Räcke [36] that describes the quality of decomposition one can achieve when decomposing the graph $G$ into a family $\mathcal{G}_T$ of its decomposition trees (cf. [36] for a formal definition of a decomposition tree).

**Theorem 3.5** ([36]). *For any graph $G = (V, E, u)$, an $\widetilde{O}(m)$-sparse $(O(\log n), \mathcal{G}_T)$-decomposition of $G$ can be found in polynomial time, where $n = |V|$ and $m = |E|$.*

Due to structural simplicity of decomposition trees, as well as, the quality of cut preservation being the best – up to a constant – achievable in this context, this theorem was used to design good approximation algorithms for a number of cut-based minimization problems.

Unfortunately, from our point of view, the usefulness of Räcke's decomposition is severely limited by the fact that the time needed to construct it is not acceptable when one is aiming at obtaining fast algorithms. More precisely, the running time of the decomposing algorithm of [36] is dominated by $\widetilde{O}(m)$ executions of the algorithm of Fakcharoenphol, Talwar and Rao [23] that solves the minimum communication cost tree problem with respect to a cost function devised from the graph $G$ (cf. [36] for details). Each such execution requires, in particular, computation of the shortest-path metric in $G$ with respect to some length function, which employs the well-known all-pair shortest path algorithm [25, 44, 2] running in $O(\min\{mn, n^{2.376}\})$ time. This results in – entirely acceptable when one is just interested in obtaining polynomial-time approximation algorithms, but prohibitive in our case – $\widetilde{O}(m \min\{mn, n^{2.376}\})$ total running time.

One could try to obtain a faster implementation of Räcke's decomposition by using a nearly-linear time low-average-stretch spanning tree construction due to Abraham, Bartal, and Neiman [1] in place of the algorithm of [23]. This would lead to an $\widetilde{O}(m^2)$ time decomposition of $G$ into its spanning trees that has a – slightly worse – quality of $\widetilde{O}(\log n)$ instead of the previous $O(\log n)$. However, although we will use [1] in our construction (cf. section 5.2), $\widetilde{O}(m^2)$ time is still not sufficiently fast for our purposes.

Our key idea for circumventing this running time bottleneck is allowing ourselves more flexibility in the choice of the family of graphs into which we will decompose $G$. Namely, we will be considering $(\alpha, \mathcal{G})$-decompositions of $G$ into objects that are still structurally simpler than $G$, but not as simple as trees.

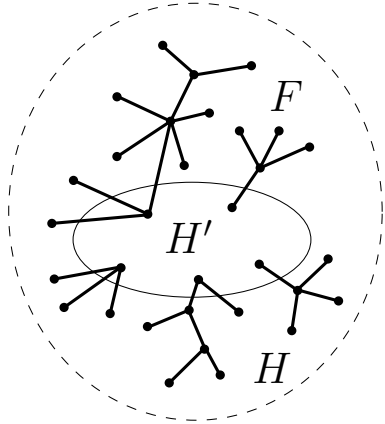

Figure 2: An example of a $j$-tree $H$, its core $H'$, and its envelope $F$ consisting of bold edges.

To this end, for $j \geq 1$, we say that a graph $H = (V_H, E_H, u_H)$ is a $j$-tree (cf. Figure 2) if it is a connected graph being a union of: a subgraph $H'$ of $H$ induced by some vertex set $V'_H \subseteq V_H$ with $|V'_H| \leq j$; and of a forest $F$ on $V_H$ whose each connected component has exactly one vertex in $V'_H$. We will call the subgraph $H'$ the *core* of $H$ and the forest $F$ the *envelope* of $H$.[6]

Now, if we define $\mathcal{G}_V[j]$ to be the family of all the $j$-trees on the vertex set $V$, the following theorem – being the heart of our framework – holds. Its proof appears in section 5.

**Theorem 3.6.** *For any graph $G = (V, E, u)$ and $t \geq 1$, we can find in time $\widetilde{O}(tm)$ a $t$-sparse $(\widetilde{O}(\log n), \mathcal{G}_V[\widetilde{O}(\frac{m \log U}{t})])$-decomposition $\{(\lambda_i, G_i)\}_i$ of $G$, where $m = |E|$, $n = |V|$, and $U$ is the capacity ratio of $G$. Moreover, the capacity ratio of each $G_i$ is $O(mU)$.*

Intuitively, the above theorem shows that if we allow $j = \widetilde{O}(\frac{m \log U}{t})$ to grow, the sparsity of the corresponding decomposition of $G$ into $j$-trees – and thus the time needed to compute it – will decrease proportionally.[7]

Note that a 1-tree is just an ordinary tree, so by taking $t$ in the above theorem sufficiently large, we obtain a decomposition of $G$ into trees in time $\widetilde{O}(m^2 \log U)$.[8] Therefore, we see that compared to the decomposition result of Räcke (cf. Theorem 3.5), we obtain in this case a decomposition of $G$ with more efficient implementation and into objects that are even simpler than decomposition trees, but at a cost of slightly worse quality.

---

[6]Note that given a $j$-tree we can find its envelope in linear time. Therefore, throughout the paper we always assume that the $j$-trees we are dealing with have their envelopes and cores explicitly marked.

[7]Interestingly, such a trade-off is somewhat reminiscent of the trade-off achieved by Spielman and Teng [40, 41] between the number of additional edges of an ultrasparsifier and the quality of spectral approximation provided by it.

[8]In fact, one might show that if our goal is to decompose $G$ into trees then the running time of our algorithm is just $\widetilde{O}(m^2)$.

However, the aspect of this theorem that we will find most useful is the above-mentioned trade-off between the simplicity of the $j$-trees into which the decomposition decomposes $G$ and the time needed to compute it. This flexibility in the choice of $t$ can be utilized in various ways.

For instance, one should note that, in some sense, the core of a $j$-tree $H$ captures all the non-trivial cut-structure of $H$. In particular, it is easy to see that maximum flow computations in $H$ essentially reduce to maximum flow computations in $H$'s core. So, one might hope that for some cut-based problems the complexity of solving them in $H$ is proportional to the complexity of solving them in the – possibly much smaller than $H$ – core of $H$ (this is, for example, indeed the case for the balanced separator and the sparsest cut problems – see section 4.3). Thus one could use Theorem 3.6 – for appropriate choice of $t$ – to get a faster algorithm for this kind of problems by just computing first the corresponding decomposition and then leveraging the existing algorithms to solve the given problem on a small number of sampled $j$-trees, while paying an additional $\widetilde{O}(\log n)$ factor in the approximation quality for this speed-up.

Even more importantly, our ability to choose in Theorem 3.6 sufficiently small value of $t$, as well as, the sparsification technique (cf. Theorem 2.3) and recursive application of the theorem to the cores of $\widetilde{O}(\frac{m \log U}{t})$-trees sampled from the computed decomposition, allows establishing the following theorem – its proof appears in section 6.

**Theorem 3.7.** *For any $1 \geq l \geq 0$, integral $k \geq 1$, and any graph $G = (V, E, u)$, we can find in $\widetilde{O}(m + 2^k n^{(1 + \frac{1-l}{2^k - 1})} \log U)$ time a collection of $(2^{k+1} \ln n)$ $n^l$-trees $\{G_i\}_i$ that $(\log^{(1+o(1))k} n)$-preserve the cuts of $G$ with high probability. Moreover, the capacity ratio of each $G_i$ is $n^{(2+o(1))k} U$, where $U$ is the capacity ratio of $G$, $n = |V|$, and $m = |E|$.*

As one can see, the above theorem allows obtaining a collection of $j$-trees that $\alpha$-preserve cuts of $G$ – for *arbitrary* $j = n^l$ – in time *arbitrarily* close to nearly-linear, but at a price of $\alpha$ growing accordingly as these two parameters decrease.

Note that one can get a cut-preserving collection satisfying the requirements of the theorem by just finding an $(\widetilde{O}(\log n), n^l)$-decomposition of $G$ via Theorem 3.6 and sampling – in the spirit of Fact 3.4 – $O(\log n)$ $n^l$-trees from it so as to ensure that each cut is preserved with high probability. Unfortunately, the running time of such procedure would be too large. Therefore, our approach to establishing Theorem 3.7 can be heuristically viewed as an algorithm that in case when the time required by Theorem 3.6 to compute a decomposition of $G$ into $n^l$-trees is not acceptable, does not try to compute and sample from such decomposition directly. Instead, it performs its sampling by finding a decomposition of $G$ into $j$-trees, for some value of $j$ bigger than $n^l$, then samples a $j$-tree from it and recurses on this sample. Now, the desired collection of $n^l$-trees is obtained by repeating this sampling procedure enough times to make sure that the cut preserving probability is high enough, with the value of $j$ chosen so as to bound the number of recursive calls by $k-1$. Since each recursive call introduces an additional $\widetilde{O}(\log n)$ distortion in the faithfulness of the reflection of the cut structure of $G$, the quality of the cut preservation of the collection generated via such algorithm will be bounded by $\log^{(1+o(1))k} n$.

## 4  Applications

We proceed to demonstrating how the tools and ideas presented in the previous section lead to a fast approximation algorithms for cut-based graph problems. In particular, we apply our techniques to the (generalized) sparsest cut and the balanced separator problems. Later we remark on the fact that our framework is also applicable to multicut-based minimization problems.

The following theorem encapsulates our way of employing our framework.

**Theorem 4.1.** *For any $\alpha \geq 1$, integral $k \geq 1$, $1 \geq l \geq 0$, and undirected graph $G = (V, E, u)$ with $n = |V|$, $m = |E|$, and $U$ being its capacity ratio, we can find in $\widetilde{O}(m + 2^k n^{(1+\frac{1-l}{2^k-1})} \log U)$ time, a collection of $2^{k+1} \ln n$ $n^l$-trees $\{G_i\}_i$, such that for* any *instance $P$ of* any *cut-based minimization problem $\mathcal{P}$ the following holds with high probability. If $\{C_i^*\}_i$ is a collection of cuts of $G$ with each $C_i^*$ being some $\alpha$-optimal solution to $P$ on the $n^l$-tree $G_i$, then at least one of $C_i^*$ is an $(\alpha \log^{(1+o(1))k} n)$-optimal solution to $P$ on the graph $G$.*

When we take $l$ equal to zero, the above theorem implies that if we have an $\alpha$-approximation algorithm for a given cut-based problem on trees that runs in $T(m,n,U)$ time then, for any $k \geq 1$, we can get an $(\alpha \log^{(1+o(1))k} n)$-approximation algorithm for it in general graphs and the running time of this algorithm will be just $\widetilde{O}(m + k2^k n^{(1+1/(2^k-1))} \log U) + (2^{k+1} \ln n)T(m,n,U)$. Note that the computational overhead introduced by our framework, as $k$ grows, quickly approaches nearly-linear. Therefore, if we are interested in designing fast poly-logaritmic approximation algorithms for some cut-based minimization problem, we can just focus our attention on finding a fast approximation algorithm for its tree instances.

Also, an interesting feature of our theorem is that the procedure producing the graphs $\{G_i\}_i$ is completely oblivious to the cut-based problem we want to solve – the fact that this problem is a cut-based minimization problem is all we need to make our approach work.

*Proof of Theorem 4.1:* We produce the desired collection $\{G_i\}_i$ by just employing Theorem 3.7 with the desired value of $k \geq 1$ to find in time $\widetilde{O}(m + 2^k n^{(1+\frac{1-l}{2^k-1})} \log U)$ a set of $t = (2^{k+1} \ln n)$ $n^l$-trees that $(\log^{(1+o(1))k} n)$-preserve the cuts of $G$ with high probability and output them as $\{G_i\}_i$.

Now, to prove the theorem, let $C^*$ be the cut that minimizes the quantity $u(C_i^*)f_P(C_i^*)$ among all the $t$ solutions $C_1^*, \ldots, C_t^*$ found, where $f_P$ is the function corresponding to the instance $P$ of a problem $\mathcal{P}$ we are solving (cf. equation (1)). Clearly, by definition of $(\log^{(1+o(1))k} n)$-preservation of the cuts, the capacity of $C^*$ – and thus the quality of the solution corresponding to it – can only improve in $G$ i.e.

$$u(C^*)f_P(C^*) \leq u_j(C^*)f_P(C^*),$$

where $G_j$ is the $n^l$-tree to which $C^*$ corresponds and, for any $i$, $u_i$ denotes the capacity function of $G_i$.

Moreover, if we look at an optimum solution $C^{OPT}$ to our problem in $G$ – i.e. $C^{OPT} = \arg\min_{\emptyset \neq C \subset V} u(C)f_P(C)$ – then, with high probability, in at least one of the $G_i$s, say in $G_{j'}$, $C^{OPT}$ has the capacity at most $(\log^{(1+o(1))k} n)$ times larger than its original capacity $u(C^{OPT})$ in $G$. As a result, for the $\alpha$-optimal solution $C_{j'}^*$ found by the algorithm in this $G_{j'}$ it will be the case that

$$u(C_{j'}^*)f_P(C_{j'}^*) \leq u_{j'}(C_{j'}^*)f_P(C_{j'}^*) \leq \alpha u_{j'}(C^{OPT})f_P(C^{OPT}) \leq \alpha(\log^{(1+o(1))k} n)u(C^{OPT})f_P(C^{OPT}),$$

where the first inequality follows from the fact that $\{G_i\}_i$ $(\log^{(1+o(1))k} n)$-preserve cuts of $G$.

But, by definition of $C^*$, $u(C^*)f_P(C^*) \leq u(C^*_{j'})f_P(C^*_{j'})$ and – by noting that $u(C^{OPT})f_P(C^{OPT})$ is the objective value of an optimal solution to our instance – we get that $C^*$ is indeed an $(\alpha \log^{(1+o(1))k} n)$-optimal solution to $P$ with high probability. ∎

### 4.1 Computing Maximum Concurrent Flow Rate on Trees

As Theorem 4.1 suggests, we should focus on designing fast approximation algorithms for tree instances of our problems. The basic tool we will use in this task is the ability to compute maximum concurrent flow rate on trees in nearly-linear time. The main reason why such a fast algorithm exists stems from the fact that if we have a demand graph $D = (V, E_D, d)$ and a tree $T = (V, E_T, u_T)$, there is a unique way of satisfying these demands in $T$. Namely, for each demand edge $e \in E_D$ the flow of $d(e)$ units of corresponding commodity has to be routed along the unique path $\mathsf{path}_T(e)$ joining two endpoints of $e$ in the tree $T$. As a result, we know that if we want to route in $T$ a concurrent flow of rate $\theta = 1$ then the total amount of flow flowing through a particular edge $h$ of $T$ is equal to

$$u^T[D](h) := \sum_{e \in E_D, h \in \mathsf{path}_T(e)} d(e).$$

Interestingly, we can compute $u^T[D](h)$ for all $h \in E_T$ in nearly-linear time.

**Lemma 4.2.** *For any tree $T = (V, E_T, u_t)$ and demand graph $D = (V, E_D, d)$, we can compute $u^T[D](h)$, for all $h \in E_T$, in $\widetilde{O}(|E_D| + |V|)$ time.*

*Proof:* To compute all the values $u^T[D](h)$, we adapt the approach to computing the stretch of edges outlined by Spielman-Teng in [41].

For a given tree $T' = (V', E_{T'}, u_{T'})$ and a demand graph $D' = (V', E_{D'}, d')$ corresponding to it, let us define the *size* $S_{D'}(T')$ of $T'$ (with respect to $D'$) as $S_{D'}(T') := \sum_{v \in V'}(1 + d_{D'}(v))$, where $d_{D'}(v)$ is the degree of the vertex $v$ in $D'$. Let us define a *splitter* $v^*_{D'}(T')$ of $T'$ (with respect to $D'$) to be a vertex of $T'$ such that each of the trees $T'_1, \ldots, T'_q$ obtained from $T'$ by removal of $v^*_{D'}(T')$ and all the adjacent edges has its size $S_{D'}(T'_i)$ being at most one half of the size of $T'$ i.e.

$$S_{D'}(T'_i) \leq S_{D'}(T')/2 = |E_{D'}| + |V'|/2,$$

for each $i$. It is easy to see that such a splitter $v^*(T')$ can be computed in a greedy fashion in $O(|V'|) = O(S_{D'}(T'))$ time.

Our algorithm for computing $u^T[D]$s works as follows. It starts with finding a splitter $v^* = v^*_D(T)$ of $T$. Now, let $E_0$ be the set of demand edges $e \in E_D$ such that $\mathsf{path}_T(e)$ contains $v^*$. Also, for $1 \leq i \leq q$, let $E_i \subseteq E_D$ be the set of edges $e \in D$ for which $\mathsf{path}_T(e)$ is contained entirely within $T_i$. Note that this partition of edges can be easily computed in $O(|E_D| + |V|) = O(S_{D'}(T'))$ time. Also, let us define $D_i$ to be the demand graph being the subgraph of $D$ spanned by the demand edges from $E_i$ (with demands inherited from $D$).

As a next step, our algorithm computes in a simple bottom-up fashion $u^T[D_0](h)$ for each $h \in E_T$. Then, for $1 \leq i \leq q$, it computes recursively the capacities $u^{T_i}[D_i]$ of all the edges of $T_i$ – note that, by definition of $D_i$, $u^{T_i}[D_i](h) = u^T[D_i](h)$ for any edge $h$ of $T_i$. Finally, for each $h \in E_T$, we output $u_T[D](h) = \sum_i u^T[D_i](h)$. Note that, since for $i \geq 1$, $u^T[D_i](h) \neq 0$ only if $h$ is a part of $T_i$, the sum $\sum_i u^T[D_i](h)$ has at most two non-zero components and thus all the values $u^T[D](h)$ can be computed in $O(|E_D| + |V|) = O(S_D(T))$ time. Furthermore, the fact that, by our choice of $v^*$, $S_{D_i}(T_i) \leq S_D(T)/2$ implies that the depth of recursion is at most $\log S_D(T)$ and the whole algorithm runs in $\widetilde{O}(S_D(T)) = \widetilde{O}(|E_D| + |V|)$ time, as desired. ∎

Now, the crucial thing to notice is that the best achievable flow rate $\theta^*$ of the maximum concurrent flow is equal to $\min_{h \in E_T} \frac{u_T(h)}{u^T[D](h)}$. Therefore, Lemma 4.2 implies the following corollary.

**Corollary 4.3.** *For any tree $T = (V, E_T, u_T)$ and demand graph $D = (V, E_D, d)$, we can find in $\widetilde{O}(|E_D| + |V|)$ time the optimal maximum concurrent flow rate $\theta^*$ and an edge $h^* \in E_T$ such that $\theta^* = \frac{u_T(h^*)}{u^T[D](h^*)}$.*

Note that we only compute the optimal flow rate of the concurrent flow and not the actual flows. In some sense, this is unavoidable – one can easily construct an example of maximum concurrent flow problem on tree, where the representation of any (even only approximately) optimal flow has size $\Omega(|E_D||V|)$.

## 4.2 Generalized Sparsest Cut Problem

We proceed to designing a fast approximation algorithm for the generalized sparsest cut problem. We start by noticing that for a given tree $T = (V, E_T, u_T)$, demand graph $D = (V, E_D, d)$, and an edge $h$ of this tree, the quantity $\frac{u_T(h)}{u^T[D](h)}$ is exactly the (generalized) sparsity of the cut that cuts in $T$ only the edge $h$. Therefore, Corollary 4.3 together with the fact that the sparsity of the sparsest cut is always an upper bound on the maximum flow rate achievable, gives us the following corollary.

**Corollary 4.4.** *For any given tree $T = (V, E, u)$ and demand graph $D = (V, E_D, d)$, an optimal solution to the generalized sparsest cut problem can be computed in $\widetilde{O}(|E_D| + |V|)$ time.*

Now, by applying Theorem 4.1 together with a preprocessing step of sparsification of $D$, we are able to obtain a poly-logarithmic approximation for the generalized sparsest cut problem in time close to linear.

**Theorem 4.5.** *For any graph $G = (V, E, u)$, demand graph $D = (V, E_D, d)$, and integral $k \geq 1$, there exists a Monte Carlo $\log^{(1+o(1))k} n$-approximation algorithm for generalized sparsest cut problem that runs in time $\widetilde{O}(m + |E_D| + 2^k n^{(1+1/(2^k-1))} \log U)$, where $n = |V|$, $m = |E|$, and $U$ is the capacity ratio of $G$.*

*Proof:* We start by employing Theorem 2.3 with $\delta$ equal to 1, to sparsify both $G$ – to obtain a graph $\widetilde{G} = (V, \widetilde{E}, \widetilde{u}))$ – and the demand graph $D$ – to obtain a demand graph $\widetilde{D} = (V, \widetilde{E}_D, \widetilde{d})$ – in total time of $\widetilde{O}(m + |E_D|)$. Note that computing sparsity of a cut with respect to these sparsified versions of $G$ and $D$ leads to a 4-approximate estimate of the real sparsity of that cut. Since this constant-factor error is acceptable for our purposes, we can focus on approximating the sparsest cut problem with respect to $\widetilde{G}$ and $\widetilde{D}$. To this end, we just use Theorem 4.1 together with Corollary 4.4 and, since both $|\widetilde{E}|$ and $|\widetilde{E}_D|$ have $\widetilde{O}(n)$ edges, our theorem follows. ∎

## 4.3 Balanced Separator and Sparsest Cut Problem

We turn our attention to the balanced separator problem. Analogously to the case of the generalized sparsest cut problem above, to employ our approach we need an efficient algorithm for the tree instances of the balanced separator problem. Unfortunately, although we can solve this problem on trees optimally via dynamic programming, there seems to be no algorithm that does it very efficiently – ideally, in nearly-linear time. Therefore, we circumvent this problem by settling for a fast but approximate solution.

Namely, we use the result of Sherman [37] who shows that, for any $\varepsilon > 0$, the balanced separator problem – as well as the sparsest cut problem – can be $O(\sqrt{\log n/\varepsilon})$-approximated in a graph $G$ – with $n$ vertices and $m$ edges – in time $\widetilde{O}(m+n^{3/2+\varepsilon})$. This running time corresponds to sparsifying $G$ and then using the fastest known algorithm for the maximum flow problem due to Goldberg and Rao [27] to perform maximum flow computations[9] on a sequence of $n^\varepsilon$ graphs that are derived in a certain way from the sparsified version of $G$.

Unfortunately, $\widetilde{O}(m+n^{3/2+\varepsilon})$ running time is still not sufficiently fast for our purposes. At this point, however, we recall that maximum flow computation in a $j$-tree reduces to the task of finding the maximum flow in its core. So, if we want to perform a maximum flow computation on a $j$-tree that has its core sparsified (i.e. the core has only $\widetilde{O}(j)$ edges), the real complexity of this task is proportional to $j$ as opposed to being proportional to $n$. This motivates us to obtaining an implementation of Sherman's algorithm on $j$-trees that achieves better running time. The proof of the following lemma appears in Appendix A.1.

**Lemma 4.6.** *For any $j$-tree $G = (V,E,u)$ and $\varepsilon > 0$, we can $O(\sqrt{\log n/\varepsilon})$-approximate the balanced separator and the sparsest cut problems in $\widetilde{O}(m + n^\varepsilon(n+j^{3/2}))$ time, where $m = |E|$ and $n = |V|$.*

Before we proceed further, we note the following lemma whose proof is presented in Appendix A.2.

**Lemma 4.7.** *For any graph $G = (V,E,u)$, we can find a $|V|^2$-approximation to the sparsest cut and the balanced separator problems in time $\widetilde{O}(|E|)$.*

Now, we can use Theorem 4.1 and Lemma 4.6 – for the right choice of $j = n^l$ – together with a simple preprocessing making the capacity ratio of the graphs we are dealing with polynomially bounded, to obtain the following result.

**Theorem 4.8.** *For any $\varepsilon > 0$, integral $k \geq 1$, and graph $G = (V,E,u)$, we can $(\log^{(1+o(1))(k+1/2)} n/\sqrt{\varepsilon})$-approximate the sparsest cut and the balanced separator problems in time $\widetilde{O}(m + 2^k n^{1+\frac{1}{3\cdot 2^k-1}+\varepsilon})$, where $m = |E|$ and $n = |V|$.*

*Proof:* Let us assume first that the capacity ratio of $G$ is polynomially bounded i.e. it is $n^{O(1)}$. In this case, we just use Theorem 4.1 on $G$ with $l = \frac{2^{k+1}}{3\cdot 2^k-1}$ to obtain a collection of $(2^{k+1}\ln n)$ $n^l$-trees $\{G_i\}_i$ in time
$$\widetilde{O}(m + 2^k n^{(1+\frac{1-l}{2^k-1})}) = \widetilde{O}(m + 2^k n^{(1+\frac{1}{3\cdot 2^k-1})}).$$

---

[9]Technically, when the input graph has $\widetilde{O}(n)$ edges, the algorithm of Goldberg and Rao finds in time $\widetilde{O}(n^{3/2}\log 1/\delta)$ an integral $s$-$t$ flow and a $s$-$t$ cut with values of respective throughput and capacity being within $(1+\delta)$ of each other. So, to ensure that this flow is optimal, one needs to use $\delta$ smaller than the inverse of the value of minimum $s$-$t$ cut of the graph. In principle, this value could be as large as $\Omega(nU)$ which would result in $\widetilde{O}(n^{3/2}\log U)$ running time. However, the graphs that are considered in Sherman's algorithm have the value of minimum $s$-$t$ cut always bounded by $n/2$.

Next, we compute – using the algorithm from Lemma 4.6 – $O(\sqrt{\log n/\varepsilon})$-approximately optimal solutions to our desired problem – being either the sparsest cut or the balanced separator problem – on each of $G_i$s in total time of

$$\widetilde{O}(m + (2^{k+1} \ln n)n^\varepsilon(n + n^{3l/2})) = \widetilde{O}(m + 2^k n^{1+\frac{1}{3 \cdot 2^k - 1}+\varepsilon}).$$

By Theorem 4.1 we know that choosing the best one among these solutions will give, with high probability, a $(\log^{(1+o(1))(k+1/2)} n/\sqrt{\varepsilon})$-approximately optimal solution that we are seeking.

To ensure that $G$ has its capacity ratio always polynomially bounded, we devise the following pre-processing procedure. First we use Lemma 4.7 to find a cut $C$ being $n^2$-approximation to the optimal solution of our desired problem. Let $\zeta$ be the sparsity of $C$. We remove all the edges in $G$ that have capacity smaller than $\zeta/mn^2$ and, for all the edge with capacity bigger than $\zeta n$, we trim their capacity to $\zeta n$. Clearly, the resulting graph $G'$ has its capacity ratio polynomially bounded. Also, the capacity of any cut in $G$ can only decrease in $G'$.

Now, we just run the approximation algorithm described above on $G'$ instead of $G$. Also, in the case of approximating the balanced separator problem, we set as our balance constant the value $c'' := \min\{c, c'\}$, where $c' := \min\{|C|, |\overline{C}|\}/n$ and $c$ is the balance constant of our input instance. Since the capacity ratio of $G'$ is polynomially bounded, the running time of this algorithm will be as desired. Moreover, we can assume that the $\alpha$-approximately optimal cut $C'$ output will have sparsity at most $\zeta$ in $G'$ - otherwise, we can just output $C$ as our solution. This means that $C'$ does not cut any edges with trimmed capacity and thus the capacity of $C'$ in $G$ (after putting the removed edges back) can be only by an *additive* factor of $\zeta/n^2$ larger than in $G'$. But since $\zeta/n^2$ is a lower bound on the sparsity of the optimal solution, we see that $C'$ is our desired $(\log^{(1+o(1))(k+1/2)} n/\sqrt{\varepsilon})$-approximately optimal solution. ∎

Recall that our main motivation to establishing Lemma 4.6 was our inability to solve the balanced separator problem on trees efficiently. However, the obtained trade-off between the running time of the resulting algorithm and the quality of the approximation provided for both the balanced separator and the sparsest cut problems is much better than the one we got for the generalized sparsest cut problem (cf. Theorem 4.5). This shows us that sometimes it is beneficial to take advantage of the flexibility in the choice of $l$ given by Theorem 4.1 by combining it with an existing fast approximation algorithm for our cut-based problem that allows a faster implementation on $j$-tree instances.

### 4.4 Extension to Multicut-based Problems

We briefly remark on the applicability of our framework to a class of undirected multicut-based minimization problems which generalize the undirected cut-based minimization ones we consider in the paper. To this end, let us define a problem $\mathcal{P}$ to be an *(undirected) multicut-based (minimization) problem* if its every instance $P \in \mathcal{P}$ on a graph $G = (V, E, u)$ can be cast as a task of finding a partition $\{C_i^*\}_{i=1}^{k^*}$ of vertices of $V$ that minimizes the quantity $u(\{C_i\}_{i=1}^k)f_P(\{C_i\}_{i=1}^k)$ over all partitions $\{C_i\}_{i=1}^k$ of the vertex set $V$. Here, $u(\{C_i\}_{i=1}^k)$ is the total capacity of all the edges of $G$ whose endpoints are in different $C_i$s. Once again, we require $f_P$ to be a non-negative function that can depend on $P$, but does not depend on the graph $G$. Note that in our definition we do not restrict a priori the number of sets into which $V$ is partitioned in considered partitions – this can be however specified by appropriate choice of the function $f_P$. Two important examples of problems that are captured by this definition are the multiway cut problem (cf. [20, 21, 18, 19, 29]) and the multicut problem (cf. [26]).

To see why our framework can be extended to handle multicut-based problems, note that for any partition $\{C_i\}_{i=1}^k$, we can express the total capacity of edges cut by it as a linear combination of the capacities of the cuts corresponding to each set $C_i$. Namely, $u(\{C_i\}_{i=1}^k) = \frac{1}{2}\sum_{i=1}^k u(C_i)$. Therefore, if $G$ is our input graph then any graph that approximates the cuts of $G$ also approximates its multicuts. This implies, in particular, that sparsification (cf. Theorem 2.3) preserves the multicuts up to a factor of $(1+\delta)$. Furthermore, by linearity of expectation, the analogs of Fact 3.2 and 3.4 hold for multicuts as well. One can check, however, that this is all that we need to prove an extension of Theorem 4.1 that handles minimization multicut-based problems.

## 5 Proof of Theorem 3.6

Let us fix throughout this section $t \geq 1$, the graph $G = (E, V, u)$ to be $(\widetilde{O}(\log n), \mathcal{G}_V[\widetilde{O}(\frac{m \log U}{t})])$-decomposed, $m = |E|$, $n = |V|$, and $U$ equal to the capacity ratio of $G$. Our proof of Theorem 3.6 consists of two main steps. First, we show how to quickly decompose $G$ into a $t$-sparse $(\widetilde{O}(\log n), \mathcal{H}[\widetilde{O}(\frac{m \log U}{t})])$-decomposition $\{(\lambda_i, H_i)\}_i$, where $\mathcal{H}[j]$ is a family of graphs that we will define shortly. Then, as a second step, we prove that for any graph $H \in \mathcal{H}[j]$ we can efficiently find a graph $\overline{G} \in \mathcal{G}_V[O(j)]$ (i.e. a graph $\overline{G}$ being a $O(j)$-tree) such that $H$ is embeddable into $\overline{G}$ and $\overline{G}$ is 9-embeddable into $H$. This will imply that both these graphs are equivalent (up to a constant) with respect to their cut-flow structure. As a result, if we consider a decomposition of $G$ into convex combination $\{(\lambda_i, \overline{G}_i)\}_i$, where each graph $\overline{G}_i$ is the graph from $\mathcal{G}_V[O(j)]$ equivalent – in the above sense – to the graph $H_i \in \mathcal{H}[j]$, then we will be able to show that this constitutes the $t$-sparse $(\widetilde{O}(\log n), \mathcal{G}_V[\widetilde{O}(\frac{m \log U}{t})])$-decomposition of $G$ that we are seeking.

### 5.1 Graphs $H(T, F)$ and the Family $\mathcal{H}[j]$

To define the family $\mathcal{H}[j]$, consider some spanning tree $T = (V, E_T)$ of $G$. There is a unique way of embedding $G$ into $T$. Namely, for each edge $e = (u, v)$ of $G$ we route the corresponding $u(e)$ units of flow along the unique $u$-$v$ path $\mathsf{path}_T(e)$ in $T$. This implies that if we want $G$ to be embeddable into $T$ then each edge $e$ of the tree $T$ has to have capacity of at least $u^T(e)$, where we define

$$u^T(e) := \sum_{e' \in E : e \in \mathsf{path}_T(e')} u(e').$$

Now, for a subset of edges $F \subseteq E_T$, let us define

$$E[T](F) := \{e \in E \ : \mathsf{path}_T(e) \cap F \neq \emptyset\}.$$

Finally, let $H(T, F) = (V, \overline{E}, \overline{u})$ be a graph (cf. Figure 3) with edge set $\overline{E} := E_T \cup E[T](F)$ and the capacities $\overline{u}(e)$, for $e \in \overline{E}$, being equal to:

$$\overline{u}(e) := \begin{cases} u(e) & \text{if } e \in E[T](F) \\ u^T(e) & \text{otherwise.} \end{cases}$$

In other words, $H(T, F)$ is a graph obtained by taking the forest corresponding to the tree $T$ with edges of $F$ removed and adding to it all the edges that are in the set $E[T](F)$ containing the edges of $E$ whose endpoints are in different components of this forest (note that $F \subseteq E[T](F)$).

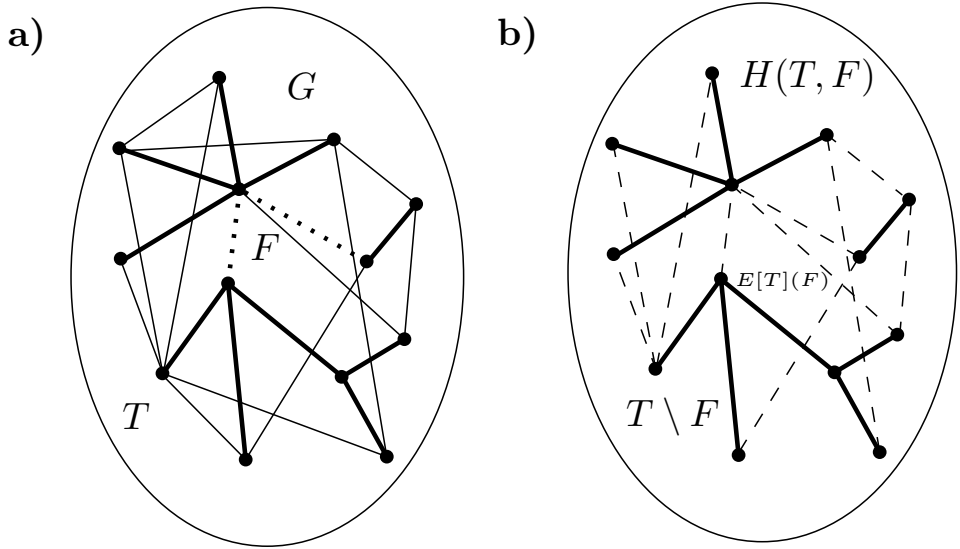

Figure 3: **a)** An example of a graph $G$ (solid edges), its spanning tree $T$ (bold edges), and a subset $F$ of edges of $T$ (dotted edges). **b)** Graph $H(T, F)$ corresponding to the example from **a)**. The edges of $T \setminus F$ are bold and the edges of the set $E[T](F)$ are dashed.

The capacity $\overline{u}(e)$ of an edge $e$ of $H(T, F)$ is equal to the capacity $u^T(e)$ inherited from the tree $T$ – if $e$ is from the forest $T \setminus F$; and it is just the original capacity $u(e)$ of $e$ otherwise.

An observation that will turn out to be useful later is that both the capacity function $u^T$ and the graph $H(T, F)$ can be constructed very efficiently – cf. Appendix B.1 for the proof of the following lemma.

**Lemma 5.1.** *Given a spanning tree $T = (V, E_T)$ of $G$ and some $F \subseteq E_T$, we can compute $u^T(e)$ for all $e \in E_T$ and construct the graph $H(T, F)$ in $\widetilde{O}(m)$ time.*

Now, we define $\mathcal{H}[j]$ to be the family of all the graphs $H(T, F)$ arising from all the possible choices of a spanning tree $T$ of $G$ and of a subset $F$ of edges of $T$ such that $|F| \leq j$.

## 5.2 Obtaining an $(\widetilde{O}(\log n), \mathcal{H}[\widetilde{O}(\frac{m \log U}{t})])$-decomposition of $G$

For a given length function $l$ on $G$ and a subgraph $H = (V, E_H, u_H)$ of $G$, let us define the *volume $l(H)$ of $H$ (with respect to $l$)* to be $l(H) := \sum_{e \in E_H} l(e) u_H(e)$. Also, for an edge $e \in E_H$, let $\gamma_H(e) := \frac{u(e)}{u_H(e)}$ and let us denote by $\gamma(H)$ the minimum value of $\gamma_H(e)$, i.e. $\gamma(H) := \min_{e \in E_H} \gamma_H(e)$. Furthermore, let us define a set $\kappa(H)$ as

$$\kappa(H) := \{e \in E_H : \gamma_H(e) \leq 2\gamma(H)\}.$$

Intuitively, $\gamma(H)$ corresponds to the inverse of maximal congestion incurred on edges of $G$ by an identity embedding of $H$ into $G$ – this identity embedding just routes the flow corresponding to given edge $e$ of $H$ along its counterpart edge in $G$. Now, $\kappa(H)$ is the set of all the edges $e$ of $H$ such that the inverse of the congestion incurred on $e$ in this identity embedding is within a factor of at most two of $\gamma(H)$.

The tool we will use to obtain an $(\widetilde{O}(\log n), \mathcal{H}[\widetilde{O}(\frac{m \log U}{t})])$-decomposition of $G$ is the following theorem that was implicitly proved by Räcke in [36] by adapting an algorithm of Young [45]. The proof of the following theorem can be found in Appendix B.2.

**Theorem 5.2.** *If there is an $\alpha \geq \ln m$ and a family of graphs $\mathcal{G}$ such that for any length function $l$ on $G$ we can find in $\widetilde{O}(m)$ time a subgraph $H_l = (V, E_{H_l}, u_{H_l})$ of $G$ that belongs to $\mathcal{G}$ and:*

(i) $l(H_l) \leq \alpha l(G)$,

(ii) $G$ is embeddable into $H_l$,

(iii) $|\kappa(H_l)| \geq \frac{4\alpha m}{t}$

*then a $t$-sparse $(2\alpha, \mathcal{G})$-decomposition of $G$ can be computed in $\widetilde{O}(tm)$ time.*

Note that in the theorem we insist that the graph $H_l$ is a subgraph of $G$. This, together with our requirement that $G$ is embeddable into $H$, implies in particular that $\gamma(H_l) \leq 1$.

At the high level, the way the decomposition is obtained in the above theorem is by iterative packing into $G$ of a $\gamma(H_l)$ fraction of the graph $H_l$ with the length functions $l$ changing from iteration to iteration. In this process, the length function models the congestion incurred on edges of $G$ by the subgraphs packed so far, i.e. the length $l(e)$ of an edge $e$ will grow exponentially with $e$'s hitherto congestion.

Recall that, by definition of $\gamma(H_l)$, we know that the packing of the $\gamma(H_l)$ fraction of the graph $H_l$ alone will not overflow the capacity of any edge of $G$. Furthermore, by insisting on each $H_l$ having relatively small volume we make sure that the whole packing does not congest any edge of $G$ by a factor greater than $\alpha$ *on average*.

Finally, by lower-bounding of the cardinality of $\kappa(H_l)$ we make sure that whenever we pack some $\gamma(H_l)$ fraction of $H_l$ we make sufficient progress – the congestion of at least $|\kappa(H_l)|$ of edges of $G$ increases by at least $1/2$. This allows us to bound the sparsity of the final decomposition, since in the end no edge has congestion bigger than $2\alpha$.

In the light of the above, we proceed to presenting an efficient construction that for any length function $l$ produces a graph $H_l \in \mathcal{H}[\widetilde{O}(\frac{m \log U}{t})]$ that satisfies the requirements of Theorem 5.2. Clearly, such a construction will immediately yield our desired $(\widetilde{O}(\log n), \mathcal{H}[\widetilde{O}(\frac{m \log U}{t})])$-decomposition of $G$.

### Construction of the Graph $H_l$

Let us fix the length function $l$ throughout this section. To explain our construction of the graph $H_l$ that satisfies the requirements of the Theorem 5.2, we first introduce the notion of low-average-stretch spanning trees – as we will see shortly such spanning trees will be the base of our construction of the graph $H_l$.

To this end, let $l'$ be some length function on a graph $G' = (V', E')$ and let us define $d_{G'}^{l'}$ to be the shortest-path metric in $G'$ with respect to this length function. Similarly, for a given spanning tree $T'$ of $G'$, let $d_{T'}^{l'}$ be the shortest-path metric in $T'$ with respect to $l'$. Now, for an edge $e = (v, w)$ of $G'$ we define the *stretch* $\mathsf{stretch}_{T'}^{l'}(e)$ of $e$ as

$$\mathsf{stretch}_{T'}^{l'}(e) := \frac{d_{T'}^{l'}(v, w)}{d_{G'}^{l'}(v, w)}.$$

The following theorem was proved by Abraham, Bartal, and Neiman [1] (see also [3] and [22] for previous results on this topic and [23] for a related result).

**Theorem 5.3** ([1]). *There is an algorithm working in $\widetilde{O}(|E'|)$ time that for any length function $l'$ on a graph $G' = (V', E')$ generates a spanning tree $T'$ of $G'$ such that the average stretch $\frac{1}{|E'|} \sum_{e \in E'} \text{stretch}_{T'}^{l'}(e)$ of edges in $T'$ is $\widetilde{O}(\log |V'|)$.*

The following lemma uses the above low-average-stretch tree construction to obtain a spanning tree $T_l$ of $G$ that – after we impose capacities $u^{T_l}$ on it – has its volume $l(T_l)$ relatively small.

**Lemma 5.4.** *We can find in $\widetilde{O}(m)$ time a spanning tree $T_l = (V, E_{T_l})$ of $G$ such that if we impose capacities $u^{T_l}$ on the edges of $T_l$ then $l(T_l) \leq 2\overline{\alpha} l(G)$, for some $\overline{\alpha}$ being $\widetilde{O}(\log n)$.*

*Proof:* Consider some spanning tree $T = (V, E_T)$ of $G$ with capacities $u^T$ on its edges.
Note that

$$l(T) = \sum_{f \in E_T} l(f) u^T(f) = \sum_{f \in E_T} l(f) \sum_{e \in E, f \in \text{path}_T(e)} u(e) = \sum_{e=(v,w) \in E} d_T^l(v,w) u(e).$$

Now, since $d_G^l(e) \leq l(e)$ for any edge $e$, we have

$$l(T) = \sum_{e=(v,w) \in E} d_{T_l}^l(v,w) u(e) \leq \sum_{e \in E} \text{stretch}_T^l(e) l(e) u(e).$$

Thus we see that to establish the lemma it is sufficient to find in $\widetilde{O}(m)$ time a tree $T$ such that

$$\sum_e \text{stretch}_T^l(e) l(e) u(e) \leq 2\overline{\alpha} l(G).$$

To this end, let us follow a technique used in [3] and define a multigraph $\overline{G}$ on vertex set $V$ that contains $r(e) := 1 + \lfloor \frac{l(e)u(e)|E|}{l(G)} \rfloor$ copies of each edge $e \in E$. Note that the total number of edges of $\overline{G}$ – when counting multiple copies separately – is

$$\sum_e r(e) \leq |E| + \sum_e \frac{l(e)u(e)|E|}{l(G)} \leq 2|E|.$$

Now, we use the algorithm[10] from Theorem 5.3 on $\overline{G}$ with length function $l$ and obtain in $\widetilde{O}(\sum_e r(e)) = \widetilde{O}(m)$ time a spanning tree $T_l$ of $\overline{G}$ – that also must be a spanning tree of $G$ – such that:

$$\frac{\sum_e \text{stretch}_{T_l}^l(e) r(e)}{\sum_e r(e)} \leq \overline{\alpha},$$

for an $\overline{\alpha}$ being $\widetilde{O}(\log n)$.
But, by the fact that for any $e$

$$r(e) \geq \frac{l(e)u(e)|E|}{l(G)} \geq \frac{l(e)u(e) \sum_{e'} r(e')}{2l(G)},$$

---

[10]Technically, the algorithm of [1] is designed for simple graphs, but it is straight-forward to adapt it to work on multigraphs with a running time being nearly-linear in the total number of edges.

we get

$$\frac{\sum_e \mathsf{stretch}^l_{T_l}(e)l(e)u(e)}{2l(G)} \leq \frac{\sum_e \mathsf{stretch}^l_{T_l}(e)r(e)}{\sum_{e'} r(e')} \leq \overline{\alpha}.$$

This means that

$$l(T_l) \leq \sum_e \mathsf{stretch}^l_{T_l}(e)l(e)u(e) \leq 2\overline{\alpha}l(G)$$

with $\overline{\alpha} = \widetilde{O}(\log n)$, as desired. ∎

The crucial fact to note at this point is that the definition of $u^{T_l}$ ensures that $G$ is embeddable into $T_l$ and thus the tree $T_l$ is guaranteed to satisfy all the requirements of Theorem 5.2 – with $\alpha$ equal to $2\overline{\alpha}$ – except condition (iii). To illustrate our way of alleviating with this shortcoming of $T_l$, consider a hypothetical situation in which $\kappa(T_l) = \{e\}$ for some edge $e$ and for all the other edges $f \neq e$ of $T_l$ we have $\gamma_{T_l}(f) = \gamma$ for some $\gamma \gg \gamma(T_l) = \gamma_{T_l}(e)$. Clearly, in this case $T_l$ is a bad candidate for the graph $H_l$ (at least when $t$ is not very large).

However, the key thing to notice in this situation is that if – instead of $H_l = T_l$ – we consider the graph $H_l = H(T_l, F)$ with $F = \{e\}$ then for all the edges $f$ of $T_l$ other than $e$ we have $\gamma_{H_l}(f) = \gamma$. Furthermore, for each edge $f \in E[T_l](\{e\})$ it is the case that $\gamma_{H_l}(f) = \frac{u(f)}{u(f)} = 1 \geq \gamma$. This means that $\gamma(H_l) = \gamma$ and thus $|\kappa(H_l)| \geq n - 2$, which makes $H_l$ satisfy condition (iii) for even very small values of $t$.

We see, therefore, that in this case by adding only one bottlenecking edge $e$ to $F$ and considering the graph $H_l = H(T_l, F)$ – as opposed to $H_l = T_l$ – we managed to make the size of the set $\kappa(H_l)$ really large. It turns out that by utilizing the fact that $1 \geq \gamma_{T_l}(e) \geq 1/mU$ and thus there is at most $\lceil \log mU \rceil$ different values of $\gamma_{T_l}(e)$ such that no two of them is within a factor of two of one another, we can always make the above approach work. Namely, as we will see in the following lemma, we can get the size of $\kappa(H_l)$ to be at least $\frac{4(2\overline{\alpha}+1)m}{t}$ while fixing in the above manner only $\widetilde{O}(\frac{m \log U}{t})$ edges of the tree $T_l$.

**Lemma 5.5.** *We can construct in $\widetilde{O}(m)$ time a graph $H_l = H(T_l, F_l)$, for some $F_l \subseteq E_{T_l}$ with $|F_l| = \widetilde{O}(\frac{m \log U}{t})$, that satisfies all the requirement of Theorem 5.2 with $\alpha$ equal to $(2\overline{\alpha}+1)$.*

*Proof:* First, we prove that no matter what is our choice of the set $F_l$ the graph $H_l = H(T_l, F_l)$ satisfies condition (i) and (ii) for $\alpha = (2\overline{\alpha}+1)$. To this end, note that

$$l(H_l) = \sum_{e \in (E_{T_l} \setminus F_l)} l(e)u^{T_l}(e) + \sum_{e \in E[T_l](F_l)} l(e)u(e) \leq l(T_l) + l(G) \leq (2\overline{\alpha}+1)l(G).$$

So, the condition (i) holds. Furthermore, it is easy to see that $G$ is embeddable into $H_l$ i.e. that the condition (ii) holds as well. We can just embed each $e \in E[T_l](F_l)$ by routing corresponding flow along the same edge $e$ in $H_l$ and each $e \notin E[T_l](F_l)$ is embed by routing $u(e)$ units of flow along the path $\mathsf{path}_{T_l}(e)$ that – by definition of $E[T_l](F_l)$ – is contained entirely in $E_{T_l} \setminus F_l$.

We proceed to finding the subset $F_l$ of edges of the tree $T_l$ that will make $H_l = H(T_l, F_l)$ satisfy also the condition (iii). Let us define for $0 \leq j \leq \lfloor \log mU \rfloor$ $F_j(T_l)$ to be the set of edges $e$ of $T_l$ such that $\frac{2^j}{mU} \leq \gamma_{T_j}(e) < \frac{2^{j+1}}{mU}$.

Note that for any edge $e$ of $T_l$ $\frac{1}{mU} \leq \frac{u(e)}{u^{T_l}(e)} \leq 1$ – the first inequality follows since in the worst case all $m$ edges will be routed in $T_l$ over $e$ and the second one since $e$ itself has to be routed in $T_l$ over $e$. Therefore, the union $\bigcup_{j=0}^{\lfloor \log mU \rfloor} F_j(T_l)$ of all the $F_j(T_l)$ partitions the whole set $E_{T_l}$ of the edges of the tree $T_l$.

Now, let us take $j^*$ to be the largest $j$ such that

$$\sum_{j'=0}^{j} |F_{j'}(T_l)| \leq \frac{4(2\overline{\alpha}+1)m(\lfloor \log mU \rfloor + 1)}{t}.$$

In other words, $j^*$ is the largest $j$ such that we could afford to take as $F_l$ the union of all the sets $F_0(T_l), \ldots, F_{j^*}(T_l)$ and still have the size of $F_l$ not exceed the desired cardinality bound of

$$\frac{4(2\overline{\alpha}+1)m(\lfloor \log mU \rfloor + 1)}{t} = \widetilde{O}(\frac{m \log U}{t}).$$

Note that we can assume that $j^* < \lfloor \log mU \rfloor$. Otherwise, we could just afford to take $F_l$ to be the union of all the sets $F_j(T_l)$ i.e. $F_l = E_{T_l}$. This would mean that $H_l = H(T_l, E_{T_l})$ is just the graph $G$ itself and thus $\gamma(H_l) = 1$ and $|\kappa(H_l)| = m$, which would satisfy the condition (iii).

Once we know that $j^* < \lfloor \log mU \rfloor$, the definition of $j^*$ implies that

$$\sum_{j'=0}^{j^*+1} |F_{j'}(T_l)| > \frac{4(2\overline{\alpha}+1)m(\lfloor \log mU \rfloor + 1)}{t}.$$

However, since this sum has $j^* + 2 \leq \lfloor \log mU \rfloor + 1$ summands, pigeon-hole principle asserts existence of $0 \leq \overline{j} \leq j^*$ such that

$$|F_{\overline{j}+1}(T_l)| \geq \frac{4(2\overline{\alpha}+1)m}{t}.$$

Now, we define $F_l$ to be the union $\bigcup_{j=0}^{\overline{j}} F_j(T_l)$. By the fact that $\overline{j} \leq j^*$ we know that the size of such $F_l$ obeys the desired cardinality bound. Furthermore, we claim that if we take $H_l = H(T_l, F_l)$ with such choice of $F_l$ then $|\kappa(H_l)|$ is large enough.

To see this, note that the fact that $\gamma_{H_l}(e) = 1$ for all edges $e$ in $E[T_l](F_l)$ implies that $\gamma(H_l)$ is at least $\frac{2^{\overline{j}+1}}{mU}$. But this means that all the edges from $F_{\overline{j}+1}(T_l)$ are in $\kappa(H_l)$ and thus

$$|\kappa(H_l)| \geq |F_{\overline{j}+1}(T_l)| \geq \frac{4(2\overline{\alpha}+1)m}{t},$$

as desired.

Now, to conclude the proof, we notice that Lemma 5.4 and Lemma 5.1 imply that the graph $H_l$ as above can indeed be constructed in $\widetilde{O}(m)$ time – to find the set $F_l$ we just sort the edges of $T_l$ according to $\gamma_{T_l}(e)$ and choose the appropriate value of $\overline{j}$. ∎

By combining Theorem 5.2, Lemma 5.5, and the fact that – by definition – $H_l \in \mathcal{H}[\widetilde{O}(\frac{m \log U}{t})]$, we get the following corollary.

**Corollary 5.6.** *A $t$-sparse $(\overline{\alpha}', \mathcal{H}[\widetilde{O}(\frac{m \log U}{t})])$-decomposition of $G$ can be found in $\widetilde{O}(tm)$ time for some $\overline{\alpha}'$ being $\widetilde{O}(\log n)$.*

## 5.3 Obtaining an $(\widetilde{O}(\log n), \mathcal{G}_V[\widetilde{O}(\frac{m \log U}{t})])$-decomposition of $G$

Let us call $G'$ an *almost-$j$-tree* if it is a union of a tree and of an arbitrary graph on at most $j$ vertices. Now, let us consider a graph $H(T, F)$ for some choice of the spanning tree $T$ of $G$ and of a subset $F$ of edges of $T$. A property of this graph that we will find useful is that the cut-flow structure of $H(T, F)$ is similar to a cut-flow structure of an almost-$O(|F|)$-tree $G(T, F)$ and this $G(T, F)$ can be found efficiently.

**Lemma 5.7.** *For any spanning tree $T = (V, E_T)$ of $G$ and $F \subseteq E_T$, we can find in $\widetilde{O}(m)$ time an almost-$O(|F|)$-tree $G(T, F)$ such that the graph $H(T, F)$ is embeddable into $G(T, F)$ and $G(T, F)$ is efficiently 3-embeddable into $H(T, F)$. Furthermore, the capacity ratio of $G(T, F)$ is at most $mU$.*

*Proof:* Consider some edge $e = (v, v')$ from $E[T](F)$. Let us define $v^1(e)$ (resp. $v^2(e)$) to be the vertex that corresponds to the first (resp. last) moment we encounter an endpoint of an edge from $F$ while moving along the path $\mathsf{path}_T(e)$ from $v$ to $v'$. Also, let us denote by $\mathsf{path}_T^1(e)$ (resp. $\mathsf{path}_T^2(e)$) the fragment of $\mathsf{path}_T(e)$ between $v$ and $v^1(e)$ (resp. between $v'$ and $v^2(e)$).

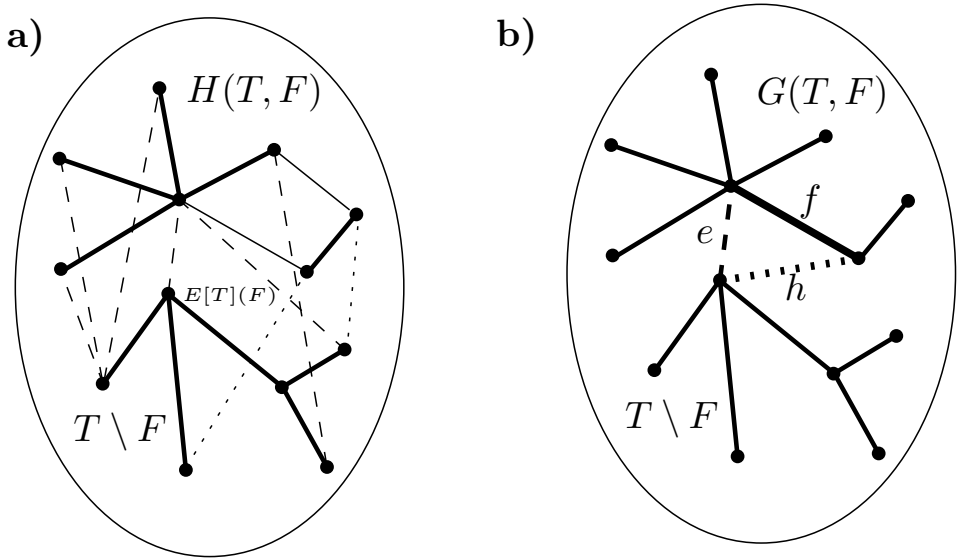

Figure 4: **a)** An example of a graph $H(T, F)$ from Figure 3. **b)** The corresponding graph $G(T, F)$ with three projected edges $e$, $f$, and $h$. The set $\mathsf{Proj}(e)$ consist of edges of $H(T, F)$ marked as dashed, the edges of the set $\mathsf{Proj}(f)$ are solid, and $\mathsf{Proj}(h)$ contains all the edges that are dotted.

We define the graph $G(T, F) = (V, E', u')$ as follows. We make the edge set $E'$ of $G(T, F)$ to consist of the edges from $E_T \setminus F$ and of an edge $f = (w, w')$ for each $w, w'$ such that $(w, w') = (v^1(e), v^2(e))$ for some edge $e \in E[T](F)$. We will call such edge $f$ *projected* and define $\mathsf{Proj}(f)$ to be the set of all $e \in E[T](F)$ with $(v^1(e), v^2(e)) = (w, w')$. Now, we set the capacity $u'(e)$ in $G(T, F)$ to be:

$$u'(e) := \begin{cases} 2u^T(e) & \text{if } e \in E_T \setminus F; \\ \sum_{e' \in \mathsf{Proj}(e)} u(e') & \text{otherwise (i.e. if } e \text{ is projected).} \end{cases}$$

See Figure 4 for an example of a graph $H(T,F)$ and the graph $G(T,F)$ corresponding to it.

To see that this construction of $G(T,F)$ can be performed quickly, note that by Lemma 5.1 we can find the set $E[T](F)$ and capacity function $u^T$ in $\widetilde{O}(m)$ time. Moreover, by employing a simple adaptation of the divide-and-conquer approach that was used in Lemma 4.2 to compute $u^T[D]$, we can compute all $v^1(e), v^2(e)$ for each $e \in E[T](F)$ also in $\widetilde{O}(m)$ time.

By noting that if a vertex $v$ is equal to $v^1(e)$ or $v^2(e)$ for some edge $e$ then $v$ has to be an endpoint of an edge in $F$, we conclude that the subgraph of $G(T,F)$ induced by projected edges is supported on at most $2|F|$ vertices. This means that $G(T,F)$ – as a connected graph being a union of a forest $E_T \setminus F$ and a graph on $2|F|$ vertices – is an almost-$2|F|$-tree. Furthermore, note that $u^T(e) \le m\max_{e'\in E} u(e')$ for each edge $e$ and that the capacity $u'(f)$ of a projected edge $f$ is upperbounded by $\sum_{e'\in E} u(e') \le m\max_{e'\in E} u(e')$. So, the fact that no edge $e$ in $G(T,F)$ has its capacity smaller than $u(e)$ implies that the capacity ratio of $G(T,F)$ is at most $mU$.

To relate the cut-flow structure of $H(T,F)$ and $G(T,F)$, one can embed $H(T,F)$ into $G(T,F)$ by embedding each edge $e \in E_T \setminus F$ of $H(T,F)$ into its counterpart edge in $G(T,F)$ and by embedding each edge $e \in E[T](F)$ by routing the corresponding flow of $u(e)$ units along the path formed from paths $\mathsf{path}_T^1(e)$ and $\mathsf{path}_T^2(e)$ connected through the projected edge $(v^1(e), v^2(e))$. It is easy to see that our definition of $u'$ ensures that this embedding does not overflow any capacities of edges of $G(T,F)$.

On the other hand, to 3-embed $G(T,F)$ into $H(T,F)$, we embed each non-projected edge of $G(T,F)$ into the same edge in $H(T,F)$. Moreover, we embed each projected edge $f$ by splitting the corresponding $u'(f) = \sum_{e\in\mathsf{Proj}(f)} u(e)$ units of flow into $|\mathsf{Proj}(f)|$ parts. Namely, for each $e \in \mathsf{Proj}(f)$, we route $u(e)$ units of this flow along the path constituted by paths $\mathsf{path}_T^1(e)$, $\mathsf{path}_T^2(e)$ and the edge $e$. Once again, it is not hard to convince oneself that such a flow does not overflow the capacities of edges of $H(T,F)$ by a factor of more than three. The lemma follows. ∎

It is easy to see that every $j$-tree is an almost-$j$-tree, however the converse does not necessarily hold – one can consider an almost-2-tree corresponding to a cycle. Fortunately, the following lemma proves that every almost-$j$-tree $G$ is close – with respect to its cut structure – to some $O(j)$-tree $\overline{G}$ and this $\overline{G}$ can be found efficiently.

**Lemma 5.8.** *Let $G' = (V', E', u')$ be an almost-$j$-tree, we can obtain in $\widetilde{O}(|E'|)$ time an $O(j)$-tree $\overline{G} = (V', \overline{E}, \overline{u})$ such that $G'$ is embeddable into $\overline{G}$ and $\overline{G}$ is 3-embeddable into $G'$. Furthermore, the capacity ratio of $\overline{G}$ is at most twice the capacity ratio of $G'$.*

*Proof:* Let us assume first that $G'$ does not have vertices of degree one - we will deal with this assumption later. Let $W \subset V'$ be the set of all vertices of degree two in $G'$. It is easy to see that we can find in $O(|E'|)$ time a collection of edge-disjoint paths $p_1, \ldots, p_k$ covering all vertices in $W$ such that in each $p_i$ all the internal vertices are from $W$ and the two endpoints $v^1(p_i)$ and $v^2(p_i)$ of $p_i$ are not in $W$, i.e. they have degree at least three in $G'$.

We construct $\overline{G}$ by first taking the subgraph $H'$ of $G'$ induced by the vertex set $V' \setminus W$ and an empty forest $\overline{F}$. Next, for each path $p_i$, we repeat the following. Let $e_{\min}(p_i)$ be the edge of $p_i$ that has minimal capacity $u'(e)$ among all the edges $e$ of $p_i$. We add to both $\overline{G}$ and $\overline{F}$ the path $p_i$ with $e_{\min}(p_i)$ removed . We set the capacities $\overline{u}$ of these added edges to be equal to their capacity in $G'$ increased by the capacity $u'(e_{\min}(p_i))$ of $e_{\min}(p_i)$. Finally, we add an edge $(v^1(p_i), v^2(p_i))$ to $\overline{G}$ with capacity equal to $u'(e_{\min}(p_i))$. See Figure 5. Clearly, such a construction can be performed in $O(|E'|)$ time and the capacity ratio of $\overline{G}$ can be at most twice the capacity ratio of $G'$.

We claim that the graph $\overline{G}$ obtained above is a $(3j-2)$-tree with $\overline{F}$ being its envelope and its core being the subgraph of $\overline{G}$ induced by vertex set $V' \setminus W$ (note that this subgraph consists of $H'$ and all the edges $\{(v^1(p_i), v^2(p_i))\}_i)$. One can see that we only need to argue that $|V' \setminus W| \le 3j+2$, the rest of the claim follows immediately from the above construction. To this end, note that since $G'$ is an almost-$j$-tree at least $|V'| - j$ of its vertices is incident only to edges of the underlying spanning tree on $V'$. This means that the total sum of degrees of these vertices in $G'$ is at most $2(|V'|-1)$. But there is no vertices in $G'$ with degree smaller than two, thus a simple calculation shows that at most $2(j-1)$ of them can have degree bigger than two. We can conclude therefore that $|W| \ge |V'| - j - 2(j-1)$ and $|V' \setminus W| \le 3j - 2$, as desired.

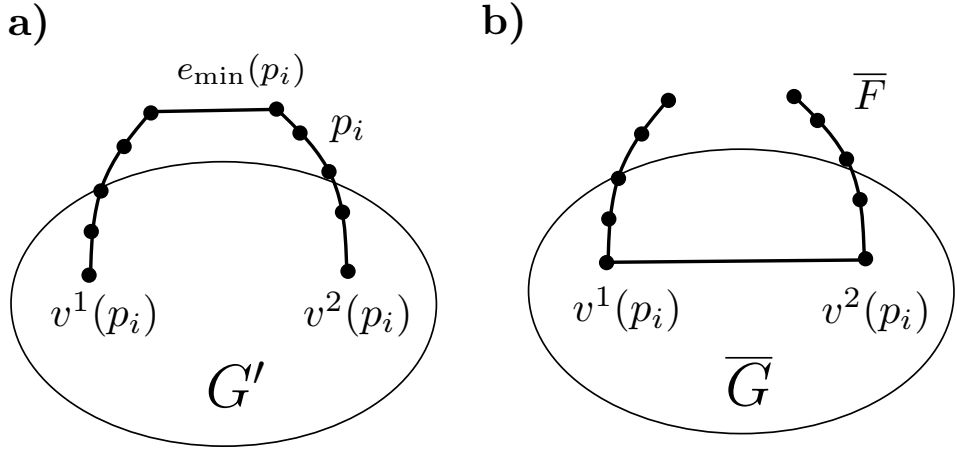

Figure 5: **a)** An example of a path $p_i$ found in $G'$. **b)** Transformed path $p_i$ in the resulting graph $\overline{G}$.

Now, to see that $G'$ embeds into $\overline{G}$, we note that $H'$ is already contained in $\overline{G}$. So, we only need to take care of the paths $\{p_i\}_i$. For each path $p_i$, $\overline{G}$ already contains all its edges except $e_{\min}(p_i)$. But this edge can be embedded into $\overline{G}$ by just routing the corresponding flow from one of its endpoints, along the fragment of path leading to $v^1(p_i)$, then through the edge $(v^1(p_i), v^2(p_i))$ and finally back along the path to the other endpoint. It is easy to verify that our way of setting up capacities ensures that this routing will not overflow any of them.

Similarly, to 3-embed $\overline{G}$ into $G'$ we first embed $H'$ into $G'$ via identity embedding. Subsequently, for each path $p_i$, $G'$ already contains all its edges, so we can use identity embedding for them as well. Furthermore, to embed the remaining edges $\{(v^1(p_i), v^2(p_i))\}_i$ we can just route, for each $i$, the corresponding flow from $v^1(p_i)$ to $v^2(p_i)$ along the path $p_i$. Once again, one can convince oneself that by definition of $e_{\min}(p_i)$ the resulting embedding does not overflow any capacities in $G'$ by a factor of more than three.

To conclude the proof, it remains to explain how to deal with graph $G'$ having vertices of degree one. In this case one may preprocess $G'$ as follows. We start with an empty forest $\overline{F}'$ and as long as there is a degree one vertex in $G'$ we remove it – together with the incident edge – from $G'$ and we add both the vertex and the edge to $\overline{F}'$. Once this procedure finishes the resulting graph $G'$ will not have vertices of degree one and still will be an almost-$j$-tree. Thus, we can use our algorithm described above and then just add $\overline{F}'$ to the $(3j-2)$-tree that the algorithm outputs. It is easy to see that the resulting graph will be still a $(3j-2)$-tree – with $\overline{F}'$ being part of its envelope – having all the desired properties. ∎

Note that the above lemmas can be combined to produce – in $\widetilde{O}(m)$ time – for a given graph $H(T, F)$ an $O(|F|)$-tree that preserves the cut-flow structure of $H(T, F)$ up to a factor of nine. This allows us to prove Theorem 3.6.

*Proof of Theorem 3.6:* By Corollary 5.6 we can compute in $\widetilde{O}(tm)$ time a $t$-sparse $(\overline{\alpha}', \mathcal{H}[\widetilde{O}(\frac{m \log U}{t})])$-decomposition $\{(\lambda_i, H(T_i, F_i))\}_i$ of $G$ with $\overline{\alpha}'$ being $\widetilde{O}(\log n)$. Now, consider a convex combination $\{(\lambda_i, \overline{G}_i)\}_i$ with each $\overline{G}_i$ being the $\widetilde{O}(\frac{m \log U}{t})$-tree produced by applying Lemma 5.7 and then Lemma 5.8 to the graph $H(T_i, F_i)$. Clearly, we can obtain this combination in $\widetilde{O}(tm)$ time.

We claim that this combination is a $t$-sparse $(\widetilde{O}(\log n), \mathcal{G}[\widetilde{O}(\frac{m \log U}{t})])$-decomposition of $G$. Obviously, $\sum_i \lambda_i = 1$ and $G$ is embeddable into each $\overline{G}_i$ since $G$ is embeddable into each $H(T_i, F_i)$ and each $H(T_i, F_i)$ is embeddable into corresponding $\overline{G}_i$. Similarly, by composing the 9-embedding of $\overline{G}_i$ into $H(T_i, F_i)$ and the identity embedding of $H(T_i, F_i)$ into $G$ (cf. embeddings $\{f_j\}_j$ in Definition 3.1), we get that $\{(\lambda_i, \overline{G}_i)\}_i$ satisfies all the requirements of a $t$-sparse $(9\overline{\alpha}', \mathcal{G}_V[\widetilde{O}(\frac{m \log U}{t})])$-decomposition. Furthermore, by Lemma 5.7 and Lemma 5.8 the capacity ratio of each $\overline{G}_i$ is at most $2mU$. The theorem follows. ∎

## 6 Proof of Theorem 3.7

The heart of the algorithm that will prove Theorem 3.7 is the procedure Find_Cut-preserving_Trees$(G, t, j)$ described in Figure 6. On the high level, this procedure given an input graph $G$ aims at finding a $j$-tree that $\alpha$-preserves the cuts of $G$ with some probability $p$, where both $\alpha$ and $p$ depend on the parameters $t$ and $j$. This is done by first sparsifying $G$ (cf. Theorem 2.3) to reduce the number of its edges, then using the decomposition of Theorem 3.6 to represent the resulting graph as a convex combination of $t$ simpler graphs ($j'$-trees, for some $j' \geq j$). Next, one of these simpler graphs $H$ is sampled – in the spirit of Fact 3.4 – and if $H$ is already a $j$-tree then it is returned. Otherwise, the procedure recursively finds a $j$-tree $\overline{G}'$ that approximates the cuts of the core of $H$ – thus $\overline{G}'$ is only defined on the vertex set of this core – and returns a $j$-tree that approximates the cuts of the whole $H$ – and thus of $G$ – by just adding the envelope of $H$ to $\overline{G}'$. An important detail is that whenever we recurse, we increase the value of parameter $t$ by squaring it. Intuitively, we may employ this more aggressive setting of the value of $t$ since the size of the problem – i.e. the size of the core of $H$ – decreased by a factor of at least $t$.

To justify our recursive step we prove the following lemma.

**Lemma 6.1.** *Let $H = (V, E, u)$ be a $j'$-tree and let $H' = (V', E', u)$ be its core. Also, let $\overline{G}' = (V', \overline{E}, \overline{u})$ be a $j$-tree on vertex set $V'$ that $\alpha$-preserves the cuts of $H'$ with probability $p$, for some $\alpha \geq 1$ and $p > 0$. Then the graph $\overline{G}$ being a union of $\overline{G}'$ and the envelope $F$ of $H$ is a $j$-tree on vertex set $V$ that $\alpha$-preserves the cuts of $H$ with probability $p$.*

*Proof:* Note that for any cut $\emptyset \neq C \subset V$, the capacity $u(C)$ of this cut in $H$ is equal to the capacity $u_{H'}(C)$ of this cut with respect to edges of $H'$ plus its capacity $u_F(C)$ with respect to edges of $F$. Similarly, the capacity of $C$ in $\overline{G}$ is equal to its capacity $\overline{u}(C)$ in $\overline{G}'$ plus the capacity $u_F(C)$. Therefore, the lemma follows by the fact that $\overline{G}'$ $\alpha$-preserves the cuts of $H'$ with probability $p$ and that after adding an envelope $F$ to it $\overline{G}'$ is still a $j$-tree. ∎

We proceed to analyzing the running time and the quality of cut preservation offered by the $j$-tree found by the procedure Find_Cut-preserving_Trees$(G, t, j)$. The proof of the following lemma can be found in Appendix C.1.

> **Procedure** Find_Cut-preserving_Trees($G, t, j$):
> **Input** : Graph $G = (V, E, u)$, parameters $t \geq 1$, and $j \geq 1$
> **Output**: A $j$-tree $\overline{G}$ on vertex set $V$
>
> Obtain a sparsification $\widetilde{G}$ of $G$ as in Theorem 2.3 with $\delta = 1$
> Find a $t'$-sparse $(\widetilde{O}(\log |V|), \mathcal{G}_V[\max\{\frac{|V|}{t}, 1\}])$-decomposition $\{(\lambda_i, \widetilde{G}_i)\}_i$ of $\widetilde{G}$ as in Theorem 3.6,
> where $t'$ is chosen so as to make the decomposition consist of $\max\{\frac{|V|}{t}, 1\}$-trees
> Sample a graph $H$ by choosing $\widetilde{G}_i$ with probability $\lambda_i$, as in Fact 3.4
> **if** $H$ is a $j$-tree **then**
> | $\overline{G} \leftarrow H$
> **else**
> | Let $H'$ be the core of $H$ and let $F$ be the envelope of $H$
> | Invoke Find_Cut-preserving_Trees($H', t^2, j$) to obtain a $j$-tree $\overline{G}'$ on vertex set $V' \subseteq V$ of the graph $H'$
> | Add $F$ to $\overline{G}'$ to get a $j$-tree $\overline{G}$ on vertex set $V$
> **end**
> **return** $\overline{G}$

Figure 6: Procedure Find_Cut-preserving_Trees($G, t, j$)

**Lemma 6.2.** *For a given graph $G = (V, E, u)$ and $j \geq 1$, and $t \geq 2$, Procedure Find_Cut-preserving_Trees($G, t, j$) works in $\widetilde{O}(m + tnT^2 \log U)$ time and returns a $j$-tree $\overline{G}$ that $(\log^{(1+o(1))T} n)$-preserves the cuts of $G$ with probability $(1/2)^T$, where $T \leq \max\{\lceil \log(\log_t(n/j) + 1) \rceil, 1\}$, $n = |V|$, $m = |E|$, and $U$ is the capacity ratio of $G$. Moreover, the capacity ratio of $\overline{G}$ is $n^{(2+o(1))T}U$.*

Now, proving Theorem 3.7 boils down to running the Find_Cut-preserving_Trees($G, t, j$) for the right setting of parameters and sufficiently many times, so as to boost the probability that the obtained collection of $j$-trees preserves cuts with high probability.

*Proof of Theorem 3.7:* To obtain the desired collection of $n^l$-trees, we just invoke procedure Find_Cut-preserving_Trees($G, t, j$) on $G$ ($2^{k+1} \ln n$) times with $t := n^{\frac{1-l}{2^k-1}}$. Note that for this choice of parameters we get that in the statement of Lemma 6.2 $T \leq k$.

Therefore, this lemma allows us to conclude that each invocation of the procedure takes $\widetilde{O}(n^{(1+1/k)} \log U)$ time to execute and thus we obtain ($2^{k+1} \ln n$) $n^l$-trees $\{G_i\}_i$ in total time of

$$\widetilde{O}(m + 2^k n^{1 + \frac{1-l}{2^k-1}} \log U).$$

(We used here the fact that it is sufficient to sparsify $G$ only once for all the procedure calls.)

Furthermore, Lemma 6.2 implies that the capacity ratio of the obtained trees is at most $n^{(2+o(1))k}U$ and each particular $n^l$-tree $G_i$ ($\log^{(1+o(1))k} n$)-preserves the cuts of $G$ with probability $(1/2)^k$. However, it is easy to see that this in turn means that the whole collection $\{\overline{G}_i\}_i$ ($\log^{(1+o(1))k} n$)-preserved the cuts of $G$ with probability $(1 - (1 - (1/2)^k)^{2^{k+1} \ln n}) = (1 - 1/n^2)$, as desired. The theorem follows. ∎

## Acknowledgments

We are grateful to Sanjeev Arora, Michel Goemans, Piotr Indyk, Jonathan Kelner, Shiva Kintali, Ofer Neiman, Richard Peng, Daniel Spielman, and anonymous reviewers for helpful comments and discussion.

# A  Appendix to section 4

## A.1  Proof of Lemma 4.6

*Proof:* We start by using Theorem 2.3 with $\delta = 1$ to sparsify the core of our $j$-tree $G$. This ensures that the resulting $j$-tree $\widetilde{G}$ has a core with $\widetilde{O}(j)$ edges. Also, it is easy to see that we can focus on approximating our problems in this graph since any approximation obtained for $\widetilde{G}$ leads to an approximation for $G$ that is only by at most a constant factor worse.

For given $\varepsilon > 0$, the algorithm of Sherman reduces the task of $O(\sqrt{\log n/\varepsilon})$-approximation of the balanced partition and the sparsest cut problems in the graph $\widetilde{G}$ to solving a sequence of $n^\varepsilon$ instances of the following problem[11]. We are given a graph $\widehat{G}$ being $\widetilde{G}$ with a source $s$ and sink $t$ added to it. We also add *auxiliary* edges that connect these two vertices with the original vertices corresponding to $\widetilde{G}$. The capacities of the auxiliary edges can be arbitrary – in particular, they can be zero which means that the corresponding edge is not present – but we require that the value of the $s$-$t$ cut $C = \{s\}$ is at most $n/2$ (thus the throughput of the maximum $s$-$t$ flow is also bounded by this value). The task is to find the maximum $s$-$t$ flow in the graph $\widehat{G}$.

Our goal is to show that we can solve any instance of the above problem in $\widetilde{O}(n + j^{3/2})$ time – this will imply our desired $\widetilde{O}(m + n^\varepsilon(n + j^{3/2}))$ total running time of Sherman's $O(\sqrt{\log n/\varepsilon})$-approximation algorithm and thus yield the proof of the lemma.

To this end, we design a method of fast compression of the graph $\widehat{G}$ to make it only consist of the core of $\widetilde{G}$, the source $s$ and the sink $t$, and the corresponding auxiliary edges (with modified capacities). This method will allow efficient – i.e. linear time – recovering from any maximum flow in the compressed graph a maximum flow in the original graph $\widehat{G}$. Moreover, the throughput of the maximum flow in the compressed graph will be at most $n/2$, as well. So, by running the algorithm of Goldberg and Rao [27] on this compressed graph and extending the computed maximum flow back, we will obtain a maximum flow in $\widehat{G}$ in time $\widetilde{O}(n + j^{3/2})$, as desired.

Our compression procedure works in steps – each such step is a local transformation that reduces the number of vertices and edges of the graph by at least one and can be implemented in constant time. In each transformation we consider a vertex $v$ in the current version $\widehat{G}'$ of $\widehat{G}$ such that $v$ has exactly one non-auxiliary edge $e$ incident to it and this edge is a part of an envelope of $\widetilde{G}$. As we will see, $\widehat{G}'$ will be always a subgraph of the graph $\widehat{G}$ and thus one can convince oneself that – as long as $\widehat{G}'$ still contains some edges of the envelope of $\widetilde{G}$ – we can always find a vertex $v$ as above.

Assume first that there is at most one auxiliary edge that is incident to $v$. (For the sake of the argument, let us use here a convention that this single auxiliary edge $e'$ is always present, but might have a capacity of zero.) In this case, we just contract the edge $e$ and set the new capacity of $e'$ to be the minimum of its previous capacity and the capacity of $e$. We claim now that given any maximum flow $f''$ in the graph $\widehat{G}''$ obtained through this contraction of $e$, we can extend it in constant time to a maximum flow $f'$ in the graph $\widehat{G}'$. To achieve this, we just transplant the flow of $|f''(e')|$ units that $f''$ routes over the edge $e'$ in $\widehat{G}''$, to a flow in $\widehat{G}'$ routed through the edges $e$ and $e'$. Note that by definition of the capacity of $e'$ in $\widehat{G}''$, the transplanted flow $f'$ is feasible in $\widehat{G}'$ and has the same throughput as $f''$ had in $\widehat{G}''$. Moreover, one can see that the throughput of the maximum flow in $\widehat{G}'$ can be at most the throughput of the maximum flow in $\widehat{G}''$ since any minimum $s$-$t$ cut in $\widehat{G}''$ can be easily extended to a minimum $s$-$t$ cut in $\widehat{G}'$ that has the same capacity. Thus $f'$ is indeed a maximum flow in $\widehat{G}'$.

---

[11]For our purposes, we use a slight generalization of the problem that is actually considered in [37].

Now, we deal with the case when there are two auxiliary edges $e', e''$ incident to $v$ – wlog, let us assume that the capacity $u''$ of $e''$ is not bigger than the capacity $u'$ of $e'$. To this end, we note that if we route $u''$ units of flow from $s$ to $t$ along the path consisting of the edges $e', e''$ then there still exists a maximum flow in $\widehat{G}'$ such that its flow-path decomposition contains the flow-path corresponding to our pushing of $u''$ units of flow over the edges $e', e''$. This implies that if we reduce the capacity of $e'$ and $e''$ by $u''$ – thus reducing the capacity of $e''$ to zero and removing it – and find the maximum flow $f''$ in the resulting graph, we can still recover a maximum flow in $\widehat{G}'$ by just adding the above-mentioned flow-path to $f''$. But, since in the resulting graph $v$ has at most one auxiliary edge incident to it – namely, $e'$ – we can use our transformation from the previous case to compress this graph further. This will reduce the number of vertices and edges of $\widehat{G}'$ by at least one while still being able to recover from a maximum flow in the compressed graph $\widehat{G}''$ a maximum flow in $\widehat{G}'$.

By noting that the compression procedure stops only when $\widehat{G}'$ contains no more edges of the envelope of $\widetilde{G}$ and thus the final compressed graph indeed consists of only (sparsified) core of $\widetilde{G}$, source $s$, and $t$, the lemma follows. ∎

## A.2 Proof of Lemma 4.7

*Proof:* Consider the case when we are interested in the sparsest cut problem – we will present variation for the balanced separator problem shortly. We sort the edges of $G$ non-increasingly according to their capacities. Let $e_1, \ldots, e_{|E|}$ be the resulting ordering. Let $r^*$ be the smallest $r$ such that the set $\{e_1, \ldots, e_r\}$ contains a spanning tree $T$ of $G$. It is easy to see that we can find such $r^*$ in $\widetilde{O}(|E|)$ time using union-find data structure.

Now, let us look at the cut $C$ in $G$ that cuts only the edge $e_{r^*}$ in $T$. Since no vertex of $G$ can have degree bigger than $|V|$, the sparsity of $C$ is at most $u(e_{r^*})|V|$. On the other hand, any cut in $G$ has to cut at least one edge $e_r$ with $r \leq r^*$. Therefore, we know that the sparsity of the optimal cut in $G$ has to be at least $u(e_{r^*})/|V|$. This implies that $C$ is the desired $|V|^2$-approximation of the sparsest cut of $G$.

In case of balanced separator problem, we define $r^*$ to be the smallest $r$ such that the largest connected component of the subgraph $E_r$ spanned by the edges $\{e_1, \ldots, e_r\}$ has its size bigger than $(1-c)|V|$, where $c$ is the desired balance constant of our instance of the problem. Once again, one can easily find such $r^*$ in $\widetilde{O}(|E|)$ time by employing union-find data structure. Let $F$ be the connected component of $E_{r^*}$ containing $e_{r^*}$. Note that by minimality of $r^*$, removing $e_{r^*}$ from $F$ disconnects it into two connected pieces and at least one of these pieces, say $F'$, has to have at least $(1-c)|V|/2$ and at most $(1-c)|V|$ vertices. Let $C$ be the cut corresponding to $F'$. Clearly, the sparsity of $C$ is at most $u(e_{r^*})|V|$. Furthermore, any cut $C^*$ with $\min\{|C^*|, |\overline{C^*}|\} \geq c|V|$ has to cut some edge $e_r$ with $r \geq r^*$. Therefore, the optimal solution can have sparsity at most $u(e_{r^*})/|V|$. This implies that $C$ is a $c'$-balanced separator with $c' = \min\{(1-c)/2, c\}$ and thus constitutes our desired $|V|^2$-(pseudo-)approximation for our instance of the balanced separator problem. ∎

# B Appendix to section 5

## B.1 Proof of Lemma 5.1

*Proof:* First, we show how to construct the set $E[T](F)$ in $O(m)$. We do this by first associating with each vertex $v$ a label indicating the connected component of the forest $T \setminus F$ to which $v$ belongs. This can be easily done in $O(n)$ time. Next, we start with $E' := \emptyset$ and for each $e$ edge of $G$ we check whether its both endpoints belong to the same connected component. If this is not the case, we add $e$ to $E'$. Clearly, this procedure takes $O(m)$ time and the resulting set $E'$ is equal to $E[T](F)$. Now, we obtain $H(T, F)$ by just taking the union of the edges from $E[T](F)$ – with capacities given by the capacity function $u$ – and of the forest $T \setminus F$ – with capacities given by the capacities given by capacity function $u^T$ that, as we will see shortly, can be computed in $\widetilde{O}(m)$ time.

To compute the capacity function $u^T$, we just note that for any $e \in E_T$, $u^T(e) = u^T[G](e)$, where $u^T[G](e)$ is defined in section 4.1. Thus we can use the algorithm from Lemma 4.2 to compute all the $u^T(e)$ in $\widetilde{O}(m)$ time. ■

## B.2 Proof of Theorem 5.2

*Proof:* Let $\{G_i\}_i$ be an enumeration of all the graphs from $\mathcal{G}$ into which $G$ is embeddable. Also, let us introduce a coefficient $\lambda_i \geq 0$ for each such $G_i$ – despite the fact that the number $N$ of such $G_i$'s can be very big, we will end up having only small number of $\lambda_i$'s non-zero. For a given graph $G_i = (V, E_i, u_i)$ and an edge $e \in E$, let us define $\mathsf{rload}_{G_i}(e) := \frac{u_i(e)}{u(e)} = \frac{1}{\gamma_{G_i}(e)}$, where we use a convention that $u_i(e) = 0$ if $e \notin E_i$ (recall that we require $G_i$ to be a subgraph of $G$ and thus $E_i \subseteq E$).

Following the approach from [36], let $M$ be an $|E| \times N$ matrix with $M_{e,i} := \mathsf{rload}_{G_i}(e)$. Clearly, if we consider a vector $\vec{\lambda} = (\lambda_1, \ldots, \lambda_N)$ then the inequality $M \cdot \vec{\lambda} \leq \beta \vec{1}$, for some $\beta > 0$, ensures that in the convex combination $\{(\lambda_i, G_i)\}_i$ of graphs $G_i$ we will have $\sum_i \lambda_i \mathsf{rload}_{G_i}(e) \leq \beta$ for each edge $e \in E$. This in turn means that $\sum_i \lambda_i u_i(e) \leq \beta u(e)$. So, if we consider the following set of inequalities

$$\begin{aligned} \mathsf{lmax}(M\vec{\lambda}) &\leq \beta \\ \sum_i \lambda_i &\geq 1 \\ \forall_i \lambda_i &\geq 0 \end{aligned}$$

where $\mathsf{lmax}(\vec{x}) := \ln \sum_{e \in E} \exp(x_e) \geq \max_{e \in E} x_e$, then a convex combination $\{(\lambda_i, G_i)\}_i$ corresponding to any $\vec{\lambda}$ satisfying them will constitute a $(\beta, \mathcal{G})$-decomposition of $G$ with the embeddings $f_i$'s being just the identity embeddings of $G_i$'s into $G$. Also, note that $\mathsf{lmax}(\vec{x}) \geq \ln m$ for any non-negative $\vec{x}$, thus $\beta$ has to be always at least $\ln m$.

The analysis from [36] shows that if for any length function $l$ we are always able to find a graph $G_{i(l)}$ such that $l(G_{i(l)}) \leq \alpha l(G)$ then a vector $\vec{\lambda}$ can be obtained that satisfies the above set of inequalities with $\beta$ equal to $\alpha + \log m \leq 2\alpha$.[12] The vector $\vec{\lambda}$ is found as follows. We start with $\vec{\lambda} = \vec{0}$. Then, as long as $\sum_i \lambda_i < 1$, we define a length function $l$ on $G$ by setting

---

[12] To understand how this statement follows from the analysis presented in [36], one should note that when we

$$l(e) := \frac{\exp((M\vec{\lambda})_e)}{u(e) \sum_{e' \in E} \exp((M\vec{\lambda})_{e'})}, \tag{2}$$

for each $e \in E$, and find the corresponding graph $G_{i(l)}$. Next, we increase the value of $\lambda_{i(l)}$ by $\min\{\gamma(G_{i(l)}), 1 - \sum_i \lambda_i\}$ and repeat the procedure if $\sum_i \lambda_i$ is still smaller than one.

We see therefore that if we use the above algorithm and each time, for the given length function $l$, we supply $G_{i(l)}$ corresponding to $H_l$ from the assumptions of the theorem – note that $H_l$ is in $\mathcal{G}$ and $G$ is embeddable into $H_l$, so $H_l$ is enumerated among the graphs $\{G_i\}_i$ – then we will obtain a vector $\vec{\lambda}$ that corresponds to a $(2\alpha, \mathcal{G})$-decomposition of $G$.

To bound the time taken by this algorithm, we need to bound the number of iterations – by the assumptions of the theorem, each iteration can be implemented in $\widetilde{O}(m)$ time. Also, note that the number of iterations determines the sparsity of the obtained decomposition.

To this end, let us introduce a potential function $\phi(\vec{\lambda}) := \sum_{e \in E} \sum_i \lambda_i \mathsf{rload}_{G_i}(e)$ that was used in [36]. Note that: initially $\phi(\vec{\lambda}) = \phi(\vec{0}) = 0$; the potential is only increasing throughout the algorithm; and at the end

$$\phi(\vec{\lambda}) = \sum_{e \in E} \sum_i \lambda_i \mathsf{rload}_{G_i}(e) \leq 2\alpha m,$$

since $\sum_i \lambda_i \mathsf{rload}_{G_i}(e) \leq \mathsf{lmax}(M\vec{\lambda}) \leq 2\alpha$ for every edge $e$.

However, by definition of $H_l$, whenever we increase $\lambda_{i(l)}$ by $\gamma(H_l)$ $\phi(\vec{\lambda})$ increases by at least $|\kappa(H_l)|/2$, since for each $e \in \kappa(H_l)$,

$$\gamma(H_l)\mathsf{rload}_{G_{i(l)}}(e) = \frac{\gamma(H_l)}{\gamma_{H_l}(e)} \geq \frac{1}{2}.$$

Therefore, by property (iii) of $H_l$, we have at most $t$ iterations and the theorem follows. ∎

## C  Appendix to section 6

### C.1  Proof of Lemma 6.2

*Proof:* Let us define $n_l$ to be an upper bound on the number of the vertices of the graph that we are dealing with at the beginning of the $l$-th recursive call of the procedure. Also, let $t_l$ be the value of $t$ and let $U_l$ be the upper bound on the capacity ratio of this graph at the same moment. Clearly, $n_0 = n$, $t_0 = t$, and $U_0 = U$. By solving simple recurrence we see that

$$t_l = t^{2^l} \tag{3}$$

and thus

$$n_{l+1} \leq \frac{n_l}{t_l} \leq \frac{n}{t^{2^{l+1}-1}}. \tag{4}$$

---

consider the length function defined in (2) then – using the terminology of [36] – $l(G_i) = \sum_e \mathsf{rload}_{G_i}(e)\mathsf{partial}'_e(M\vec{\lambda}) = \mathsf{partial}_i(\vec{\lambda})$ and $l(G) = 1$.

Now, by Theorem 2.3 and Theorem 3.6, we know that $U_{l+1}$ is $O(n_l^2 U_l)$. Moreover, by the fact that we sparsify $G$ first and by Theorem 3.6, we know that in $l$-th iteration it is sufficient to take $t' = t_l \hat{t}_l$, where $\hat{t}_l$ is $\log^{O(1)} n \log U_l$.

Note that every graph on at most $j$ vertices is a $j$-tree thus the procedure stops issuing further recursive calls once the number of vertices of $G$ is at most $j$. So, the number of recursive calls of our procedure is at most $T - 1$, where

$$T := \max\{\lceil \log(\log_t(n/j) + 1) \rceil, 1\}.$$

As a result, the bound on the capacity ratio of $\overline{G}$ is $U_T \leq n^{(2+o(1))T} U$, as desired.

To bound the total running time of the procedure, consider its modification in which we additionally sparsify the graph $H'$ before passing it to the recursive call. Clearly, running time of such modification can only increase. Note that in this case all but the topmost invocation of the procedure deals from the beginning till the end with sparsified version of the graph. Therefore, the time need to execute $l$-th recursive call, for $l \geq 1$, is $\widetilde{O}(\hat{t}_l t_l n_l)$. By equations (3) and (4), this implies that the total time taken by the execution of our procedure is:

$$\widetilde{O}(m) + \sum_{l=0}^{T-1} \widetilde{O}(\hat{t}_l t_l n_l) \leq \widetilde{O}(m + \sum_{l=0}^{T-1} \frac{nt^{2^l} \log U_l}{t^{2^l-1}}) \leq \widetilde{O}(m + tn \sum_{l=0}^{T-1} \log U_l) = \widetilde{O}(m + tnT^2 \log U),$$

where $\widetilde{O}(m)$ is the cost of the initial sparsification of $G$.

To establish the desired bound on the quality of the preservation of cuts of $G$ by $\overline{G}$, we prove that for all $l \geq 0$, if the execution of the procedure made $l$ recursive calls then the graph $\overline{G}$ $(\log^{(1+o(1))(l+1)} n)$-preserves the cuts of $G$ with probability $(1/2)^{l+1}$. Clearly, this claim implies the desired bounds by combining it with the fact that $l \leq T - 1$.

We prove the claim inductively. For $l = 0$ it follows from Theorem 2.3, Theorem 3.6, and Fact 3.4. Now, assume that the claim holds for $l \geq 0$ and we will prove it for $l + 1$. By Theorem 2.3, Theorem 3.6, and Fact 3.4, we know that $H$ is $(\log^{1+o(1)} n)$-preserving the cuts of $G$ with probability $1/2$. So, by our inductive assumption we know that $\overline{G}'$ $(\log^{(1+o(1))(l+1)} n)$-preserves the cuts of $H'$ with probability $(1/2)^{l+1}$. Therefore, by Lemma 6.1 we know that $\overline{G}$ also $(\log^{(1+o(1))(l+1)} n)$-preserves the cuts of $H$ with probability $(1/2)^{l+1}$ which – by the above-mention relation of the cut-structure of $H$ to the cut-structure of $G$ – implies that $\overline{G}$ $(\log^{(1+o(1))(l+2)} n)$-preserves the cuts of $H$ with probability $(1/2)^{l+2}$. So, our claim holds for $l+1$ and thus for all the values of $l$. This concludes the proof of the lemma. ∎